\documentclass[letterpaper,twocolumn,10pt]{article}
\usepackage{usenix,epsfig,endnotes}

\newif\ifshowcomments
\newif\ifanonymous

\showcommentsfalse
\anonymousfalse

\usepackage{IEEEtranstools}

\usepackage[T1]{fontenc}
\usepackage{amsmath}
\usepackage{amsfonts}
\usepackage{amsthm}
\usepackage{algorithm}
\usepackage{algorithmic}
\usepackage{xcolor}
\usepackage{tikz}
\usepackage{enumitem}

\usepackage{array}

\usepackage{pgfplots}
\usepackage{versions}
\usepackage{booktabs}
\usepackage{pdfpages}
\usepackage{url}
\usepackage{subcaption}
\usepackage[font=small,labelfont=bf]{caption}

\usepackage{xspace}
\usepackage[capitalise]{cleveref}
\includeversion{xtoc}

\newtheorem{theorem}{Theorem}
\newtheorem{lemma}{Lemma}

\usetikzlibrary{arrows}
\usetikzlibrary{intersections}
\usetikzlibrary{positioning}
\usetikzlibrary{shapes,snakes}
\usetikzlibrary{backgrounds,calc}
\usetikzlibrary{fit}
\usetikzlibrary{trees}

\tikzstyle{bag} = [text width=8em,
text centered]

\tikzstyle{bag_mod} = [text width=2em,
text centered]

\tikzstyle{bag_rect} = [draw=black,rectangle, black,text width=8em,
text centered]

\tikzstyle{bag1} = [draw=black,rectangle, black,text width=4em,
text centered]

\newcommand{\Cross}{$\mathbin{\tikz [x=1.4ex,y=1.4ex,line width=.2ex, red] \draw (0,0) -- (1,1) (0,1) -- (1,0);}$}%

\newcommand{\Checkmark}{$\color{green}\checkmark$}

\usepackage{sansmath}
\usepackage{soul}

\ifshowcomments
	\newcommand{\george}[1]{\textcolor{green}{George: #1}}
	\newcommand{\jamie}[1]{\textcolor{red}{Jamie: #1}}
	\newcommand{\ania}[1]{\textcolor{cyan}{Ania: #1}}
	\newcommand{\sebastian}[1]{\textcolor{blue}{Sebastian: #1}}
	\newcommand{\tariq}[1]{\textcolor{orange}{Tariq: #1}}
	\newcommand{\cd}[1]{\textcolor{pink}{Claudia: #1}}
\else
	\newcommand{\george}[1]{}
	\newcommand{\jamie}[1]{}
	\newcommand{\ania}[1]{}
	\newcommand{\sebastian}[1]{}
	\newcommand{\tariq}[1]{}
	\newcommand{\cd}[1]{}
\fi

\newcommand{\sysname}{Loopix\xspace}

\newcommand{\phead}[1]{\noindent{\bf #1.}}

\begin{document}

\title{\Large \bf The Loopix Anonymity System}

\ifanonymous
	\author{
	{\rm Anonymized for review} 
	}
\else
	\author{
	{\rm Ania Piotrowska} \\
	University College London
	\and
	{\rm Jamie Hayes} \\
	University College London
	\and
	{\rm Tariq Elahi} \\
	KU Leuven
	\and
	{\rm Sebastian Meiser} \\
	University College London
	 \and
	{\rm George Danezis} \\
	University College London
	}
\fi

\maketitle


\newcommand{\mathcmd}[1]{{\normalfont\ensuremath{#1}}\xspace}
\newcommand{\mathvar}[1]{\mathcmd{\mathsf{#1}}}
\newcommand{\mathstring}[1]{\mathcmd{\text{\texttt{#1}}}}
\newcommand{\mathname}[1]{\mathcmd{\text{\textit{#1}}}}
\newcommand{\mathalg}[1]{\mathcmd{\text{\textsc{#1}}}}
\newcommand{\mathtext}[1]{\mathcmd{\text{#1}}}
\newcommand{\mathvalue}[1]{\mathcmd{\mathit{#1}}}
\newcommand{\mathentity}[1]{\mathcmd{\text{\textsc{#1}}}}

\newcommand{\set}[1]{\ensuremath{\left\{#1\right\}}\xspace}
\newcommand{\pr}[1]{\ensuremath{\mathtext{Pr}\left[#1\right]}\xspace}
\newcommand{\condpr}[2]{\ensuremath{\mathtext{Pr}\left[#1 \middle| #2\right]}\xspace}
\newcommand{\from}{\ensuremath{\leftarrow}\xspace}
\newcommand{\randomfrom}{\ensuremath{\overset{R}{\leftarrow}}\xspace}

\newcommand{\adv}{\ensuremath{\mathcal A}\xspace}
\newcommand{\bdv}{\ensuremath{\mathcal B}\xspace}
\newcommand{\Chal}{\ensuremath{\mathcal C}\xspace}
\newcommand{\fetch}[1]{\ensuremath{\mathsf{fetch(#1)}}\xspace}
\newcommand{\TTP}{\ensuremath{\mathsf{TTP}}\xspace}
\newcommand{\pois}{\ensuremath{Pois}\xspace}

\newcommand{\delayT}{\mathvar{T}}
\newcommand{\delayTmax}{\mathvar{T^{max}}}
\newcommand{\delayTmin}{\mathvar{T^{min}}}
\newcommand{\delayTi}{\mathvar{T_i}}

\newcommand{\sender}[1]{$S_{#1}$\xspace}
\newcommand{\receiver}[1]{$R_{#1}$\xspace}
\newcommand{\Mixnode}[1]{$\mathsf{M_{#1}}$\xspace}
\newcommand{\mixnode}{\mathentity{M}\xspace}
\newcommand{\provider}{\mathentity{P}\xspace}
\newcommand{\packet}{\mathentity{p}\xspace}
\newcommand{\noisepacket}{\ensuremath{\packet_\mathtext{noise}}\xspace}
\newcommand{\delayqueue}{\ensuremath{\mathvar{q_{delay}}}\xspace}
\newcommand{\outputqueue}[1]{\ensuremath{\mathvar{q_{out(#1)}}}\xspace}

\newcommand{\header}[1]{$\mathsf{H_{#1}}$\xspace}
\newcommand{\MAC}[1]{$\mathsf{MAC_{#1}}$\xspace}
\newcommand{\BETA}[1]{$\mathsf{\beta_{#1}}$\xspace}
\newcommand{\ELEMENT}[1]{$\mathsf{\alpha_{#1}}$\xspace}
\newcommand{\ENC}[2]{$\mathsf{Enc_{#1}(#2)}$\xspace}
\newcommand{\FUNCTION}[2]{$\mathsf{h_{#1}(#2)}$}
\newcommand{\SECRET}[1]{$\mathsf{k_{#1}}$}
\newcommand{\msg}[1]{$\mathsf{m}$\xspace}

\subsection*{Abstract}
We present \emph{\sysname}, a low-latency anonymous communication system that provides
bi-directional `third-party' sender and receiver anonymity and
unobservability. 
\sysname leverages cover
traffic and brief message delays to provide anonymity and achieve traffic analysis resistance, including against a global network adversary. Mixes and clients self-monitor the network via loops of traffic to 
provide protection against active attacks, and inject cover traffic to provide stronger anonymity and a measure of sender and receiver 
unobservability. Service providers mediate access in and out of a stratified network of Poisson mix nodes to facilitate accounting and off-line message reception, as well as to keep the number of links in the system low, and to concentrate cover traffic. 

We provide a theoretical analysis of the Poisson mixing strategy as well as an empirical evaluation of the anonymity provided by the protocol and a functional implementation that we analyze in terms of scalability by running it on AWS EC2. We show that a \sysname relay can handle upwards of 300 messages per second, at a small delay overhead of less than $1.5\,ms$ on top of the delays introduced into messages to provide security. Overall message latency is in the order of seconds -- which is low for a mix-system. Furthermore, many mix nodes can be securely added to a stratified topology to scale throughput without sacrificing anonymity.



\section{Introduction}


In traditional communication security, the confidentiality of messages is protected through encryption, which exposes meta-data, such as who is sending messages to whom, to network eavesdroppers. As illustrated by recent leaks of extensive mass surveillance programs\footnote{See EFF's guide at \url{https://www.eff.org/files/2014/05/29/unnecessary_and_disproportionate.pdf}}, exposing such meta-data leads to significant privacy risks.



Since 2004, Tor~\cite{dingledine2004tor}, a practical manifestation of circuit-based onion routing, has become the most popular anonymous
communication tool, with systems such as Herd~\cite{le2015herd}, Riposte~\cite{corrigan2015riposte}, HORNET~\cite{ChenABDP15}
and Vuvuzela~\cite{van2015vuvuzela} extending and strengthening this circuit-based paradigm. Message-oriented architectures, based on
mix networks, have become unfashionable due to perceived higher latencies, that cannot accommodate real-time communications. However,
unless cover traffic is employed, onion routing is susceptible to traffic analysis attacks~\cite{cai2012touching} by an adversary that can monitor network links between nodes. Recent revelations suggest that capabilities of large intelligence agencies approach that of global passive observers---the most powerful form of this type of adversary.

However, it is not sufficient to provide strong anonymity against such an adversary while providing low-latency communication. A successful system additionally needs to resist powerful active attacks and use an efficient, yet secure way of transmitting messages. Moreover, the system needs to be scalable to a large number of clients, which makes classical approaches based on synchronized rounds infeasible. 

For this reason we reexamine and reinvent mix-based architectures, in the form of the \sysname anonymity system. \sysname is
resistant against powerful adversaries who are capable of observing all communications and performing active attacks.
We demonstrate that such a mix architecture can support low-latency communications that can tolerate small delays, at the cost of using some extra bandwidth for cover traffic. Delay, cover and real traffic can be flexibly traded-off against each other to offer resistance to traffic analysis.
\sysname provides \emph{`third-party'} anonymity, namely it hides the sender-receiver relationships from third parties, but senders and recipients can identify one another. This simplifies the design of the system, prevents abuse, and provides security guarantees against powerful active adversaries performing $(n-1)$ attacks~\cite{serjantov2002trickle}.

\sysname provides anonymity for private email or instant messaging applications. For this reason,
we adopt and leverage an architecture by which users of \sysname are associated with service providers that mediate their access
to a stratified anonymity system. Such providers are only semi-trusted\footnote{Details about the threat model are in \Cref{section:securityproperties}}, and are largely present to ensure messages sent to off-line users
can be retrieved at a later time, maintain
accounting, and enforce rate limiting. To provide maximal flexibility, \sysname only guarantees unreliable datagram transmission and is carried over UDP.  Reliable transport is left to the application as an end-to-end concern~\cite{saltzer1984end}.

\vspace{2mm}
\phead{Contributions} In this paper we make the following contributions:
\begin{itemize}[noitemsep, topsep=0pt]
	\item We introduce \sysname, a new message-based anonymous communication system. It allows for a tunable trade-off between latency, genuine and cover traffic volume, to foil traffic analysis.
	\item As a building block of \sysname we present the \emph{Poisson Mix}, and provide novel theorems about its properties and ways to analyze it as a pool-mix. Poisson mixing does not require synchronized rounds, can be used for low-latency anonymous communication, and provides resistance to traffic analysis. 
	\item We analyze the \sysname system against a strong, global passive adversary. Moreover, we show that \sysname provides resistance against active attacks, such as trickling and flooding. We also present a methodology to estimate empirically the security provided by particular mix topologies and other security parameters.
    \item We provide a full implementation of \sysname and measure its performance and scalability in a cloud hosting environment.
\end{itemize}

\paragraph{Outline.}
The remainder of this paper is organized as follows. In \Cref{sec:model}, we present a brief, high-level overview of \sysname and define the security goals and threat model. In \Cref{section:architecture}, we detail the design of \sysname and describe Poisson mixes, upon which \sysname is based and  introduce their properties.
In \Cref{sec:analysis}, we present the analysis of \sysname security properties and discuss the resistance against traffic analysis and active attacks.
In \Cref{sec:implementation}, we discuss the implementation of \sysname system and evaluate the performance. In \Cref{sec:related_works}, we survey related works and compare \sysname with recent designs of anonymity systems. In \Cref{sec:discussion}, we discuss remaining open problems and possible future work. Finally, we conclude in \Cref{sec:conclusion}.


\section{Model and Goals}\label{sec:model}

In this section, we first outline the design of \sysname. Then we discuss the security goals and types of adversaries which \sysname guarantees users' privacy against. 

\subsection{High-level overview}\label{section:overview}


\sysname is a mix network~\cite{Chaum81} based architecture allowing \emph{users}, distinguished as \emph{senders} and \emph{receivers},  
to route messages anonymously to each other using an infrastructure of \emph{mix} servers, acting as relays. These mix servers are arranged in a stratified topology~\cite{dingledine2004synchronous} to ensure both horizontal scalability and a sparse topology that concentrates traffic on fewer links~\cite{danezis2003mix}.
Each user is allowed to access the \sysname network through their association with a \emph{provider}, a special type of mix server.  
Each provider has a long-term relationship with its users and may authenticate them, potentially bill them or discontinue their access to the network.  
The provider not only serves as an access point, but also stores users' incoming messages. These messages can be retrieved at any time,  hence  
users do not have to worry about lost messages when they are off-line.  
In contrast to previous anonymous messaging designs~\cite{van2015vuvuzela, corrigan2015riposte}, \sysname does not operate in deterministic rounds, but runs as a continuous system. Additionally, \sysname uses the Poisson mixing technique that is based on the independent delaying of messages, which makes the timings of packets unlinkable.
This approach does not require the synchronization of client-provider rounds and does not degrade the usability of the system for temporarily off-line clients. 
Moreover, \sysname introduces different types of cover traffic to foil de-anonymization attacks. 

\subsection{Threat Model}\label{section:threadmodel}
\sysname assumes sophisticated, strategic, and well-resourced adversaries concerned with linking users to their communications and/or their communication partner(s). As such, \sysname considers adversaries with three distinct \emph{capabilities}, that are described next.

Firstly, a \emph{global passive adversary} (GPA) is able to observe all network traffic between users and providers and between mix servers. This adversary is able to observe the entire network infrastructure, launch network attacks such as BGP re-routing~\cite{ballani2007study} or conduct indirect observations such as load monitoring and off-path attacks~\cite{gilad2012spying}. Thus, the GPA is an abstraction that represents many different classes of adversaries able to observe some or all information between network nodes. 

Secondly, the adversary has the ability to observe all of the internal state of some corrupted or malicious mix relays. The adversary may inject, drop, or delay messages. She also has access to, and operates, using the secrets of those compromised parties. Furthermore, such corrupted nodes may deviate from the protocol, or inject malformed messages. A variation of this ability is where the mix relay is also the provider node meaning that the adversary additionally knows the mapping between clients and their mailboxes. We say that the provider is \emph{corrupt}, but is restricted to being honest but curious. In \sysname, we assume that a fraction of mix/provider relays can be corrupted or are operated by the adversary.

Finally, the adversary has the ability to participate in the \sysname system as a compromised user, who may deviate from the protocol. We assume that the adversary can control a limited number of such users---excluding Sybil attacks~\cite{douceur2002sybil} from the \sysname threat model---since we assume that \emph{honest providers} are able to ensure that at least a large fraction of their users base are genuine users faithfully following all \sysname protocols. Thus, the fraction of users controlled by the adversary may be capped to a small known fraction of the user base. We further assume that the adversary is able to control a compromised user in a conversation with an honest user, and become a \emph{conversation insider}.

An adversary is always assumed to have the GPA capability, but other capabilities depend on the adversary. 
We evaluate the security of \sysname in reference to these capabilities.

%

\subsection{Security Goals}\label{section:securityproperties}

The \sysname system aims to provide the following security properties against both passive and active attacks---including end-to-end correlation and $(n-1)$ attacks. These properties are inspired by the formal definitions from Anoa~\cite{backes2013anoa}.  
All security notions assume a strong adversary with information on all users, with up to one bit of uncertainty. In the following we write $\{S \rightarrow R\}$ to denote a communication from the sender $S$ to the receiver $R$, $\{S \rightarrow \}$ to denote that there is a communication from $S$ to any receiver and $\{S \not \rightarrow \}$ to denote that there is no communication from $S$ to any receiver ($S$ may still send cover messages). Analogously, we write $\{ \rightarrow R\}$ to denote that there is a communication from any sender to the receiver $R$ and $\{ \not \rightarrow R\}$ to denote that there is no communication from any sender to $R$ (however, $R$ may still receive cover messages).

 \vspace{2mm}

\phead{Sender-Receiver Third-party Unlinkability}
The senders and receivers should be unlinkable by any unauthorized party. Thus, we consider an adversary that wants to infer whether two users are communicating. We define \emph{sender-receiver third party unlinkability} as the inability of the adversary to distinguish whether $\{S_1 \rightarrow R_1,\allowbreak S_2 \rightarrow R_2\}$ or $\{S_1 \rightarrow R_2, S_2 \rightarrow R_1\}$ for any concurrently online honest senders $S_1, S_2$ and honest receivers $R_1, R_2$ of the adversary's choice.

\sysname provides strong sender-receiver third-party anonymity against the GPA even in collaboration with corrupt mix nodes.
We refer to \Cref{sec:poisson_sec} for our analysis of the unlinkability provided by individual mix nodes, to \Cref{sec:secsim} for a quantitative analysis of the sender-receiver third-party anonymity of \sysname against the GPA and honest-but-curious mix nodes and to \Cref{sec:traffic_analysis} for our discussion on active attacks.

 \vspace{2mm}

\phead{Sender online unobservability}
Whether or not senders are communicating should be hidden from an unauthorized party. We define \emph{sender online unobservability} as the inability of an adversary to decide whether a specific sender $S$ is communicating with any receiver $\{S \rightarrow\}$ or not $\{S \not \rightarrow\}$, for any concurrently online honest sender $S$ of the adversary's choice.

\sysname provides strong sender online unobservability against the GPA in collaboration with an \emph{insider} and even against a \emph{corrupt provider}. We refer to \Cref{sec:client-provider-unobservability} for our analysis of the latter.


Note, that sender online unobservability directly implies the notion of \emph{sender anonymity} where the adversary tries to distinguish between two possible senders communicating with a target receiver. Formally, $\{S_1 \rightarrow R, S_2 \not \rightarrow \}$ or $\{S_1 \not \rightarrow , S_2 \rightarrow R\}$ for any concurrently online honest senders $S_1$ and $S_2$ and any receiver of the adversary's choice. \sysname provides sender anonymity even in light of a conversation insider, i.e., against a corrupt receiver.

 
 \vspace{2mm}
 
 \phead{Receiver unobservability}
Whether or not receivers are part of a communication should be hidden from an unauthorized party. We define \emph{receiver unobservability} as the inability of an adversary to decide whether there is a communication from any sender to a specific receiver $R$ $\{\rightarrow R\}$ or not $\{\not \rightarrow R\}$, for any online or offline honest receiver $R$ of the adversary's choice.

\sysname provides strong receiver unobservability against the GPA in collaboration with an \emph{insider}, under the condition of an \emph{honest provider}. We show in \Cref{sec:client-provider-unobservability} how an honest provider assists the receiver in hiding received messages from third party observers. 

Note, that receiver unobservability directly implies the notion of \emph{receiver anonymity} where the adversary tries to distinguish between two possible receivers in communication with a target sender. Formally, $\{S \rightarrow R_1, \not \rightarrow R_2\}$ or $\{\not \rightarrow R_1, S \rightarrow R_2\}$ for any concurrently online honest sender $S$ and any two honest receivers $R_1, R_2$ of the adversary's choice.~\footnote{If the receiver's provider is honest, \sysname provides a form of receiver anonymity even in light of a conversation insider: a corrupt sender that only knows the pseudonym of a receiver cannot learn which honest client of a provider is behind the pseudonym.}

\vspace{2mm}

\phead{Non-Goals}\ 
\sysname provides anonymous unreliable datagram transmission, as well as facilities the reply of sent messages (through add-ons). This choice allows for flexible traffic management, cover traffic, and traffic shaping. On the downside, higher-level applications using \sysname need to take care of reliable end-to-end transmission and session management. We leave the detailed study of those mechanisms as future work.

The provider based architecture supported by \sysname aims to enable managed access to the network, support anonymous blacklisting to combat abuse~\cite{henry2013thinking}, and payments for differential access to the network~\cite{AndroulakiRSSB08}. However, we do not discuss these aspects of \sysname in this work, and concentrate instead on the core anonymity features and security properties described above.


\section{The Loopix Architecture}\label{section:architecture}

In this section we describe the \sysname system in detail---\Cref{fig:loopix} provides an overview. We also introduce the notation used further in the paper, summarized in \Cref{Notation}. 

\begin{table}[t!]
\centering
\small
\label{Notation}
\begin{tabular}{*2l}    \toprule
Symbol & Description  \\\midrule
$N$ & Mix nodes \\
$P$ & Providers \\
$\lambda_L$  & Loop traffic rate (user)    \\ 
$\lambda_D$  & Drop cover traffic rate (user)    \\ 
$\lambda_P$  & Payload traffic rate (user)    \\ 
$l$ & Path length (user) \\ 
$\mu$ & The mean delay at mix $M_{i}$ \\ 
$\lambda_M$  & Loop traffic rate (mix)    \\ 
\bottomrule
 \hline
\end{tabular}
\caption{Summary of notation}
\end{table}

\subsection{System Setup}\label{sec:setup} 
The \sysname system consists of a set of  mix nodes $N$ and  providers $P$. We consider a population of $U$ users communicating through \sysname, each of which can act as \emph{sender} and \emph{receiver}, denoted by indices $S_{i}$, $R_{i}$, where $i \in \{1, \ldots, U\}$ respectively. 
Each entity of the \sysname infrastructure has its unique public-private key pair $(sk, pk)$.
In order for a \emph{sender} $S_{i}$, with a key pair $(sk_{S_{i}}, pk_{S_{i}})$, to send a message to a \emph{receiver} $R_{j}$, with a key pair $(sk_{R_{j}}, pk_{R_{j}})$, the sender needs to know the receiver's \sysname \emph{network location}, i.e., the IP address of the user's provider and an identifier of the user, as well as the public encryption key $pk_{R_{j}}$.
We assume this information can be made available through a privacy-friendly lookup or introduction system for initiating secure connection~\cite{lazar2016alpenhorn}. This is out of scope for this work.

\subsection{Format, Paths and Cover Traffic}

\paragraph{Message packet format.}
All messages are \emph{end-to-end encrypted} and encapsulated into packets to be processed by the mix network. 
We use the Sphinx packet design~\cite{danezis2009sphinx}, to ensure that intermediate mixes learn no additional 
information beyond some routing information. 
All messages are padded to the same length, which hides the path length and the relay position
and guarantees unlinkability at each hop of the messages' journey over the network.
Each message wrapped into the Sphinx packet consists of two separate parts: 
a header $H$, carrying the layered encryption of meta-data for each hop, 
and the encrypted payload $\rho$ of the message. 
The header provides each mix server on the path with confidential meta-data. The meta-data  
includes a single element of a cyclic group (used to derive a shared encryption/decryption key),
the routing information and the message authentication code. We extend the Sphinx packet format to carry additional
routing information in the header to each intermediate relay, including a delay and additional flags.

\begin{center}
\begin{figure}[t!]
	\resizebox{1.\columnwidth}{!}{
	\begin{tikzpicture}[font=\sffamily]
		\node[draw, cloud, cloud puffs=12.7, minimum width = 10cm, minimum height = 10.0cm, line width = 3pt, opacity=0.5] (cloud) {};
		\node[draw=none, fill=none, above = 0.5cm of cloud] {};
		\node[draw=none, line width = 1pt] (node5)  at (cloud) {\includegraphics[width=.08\textwidth]{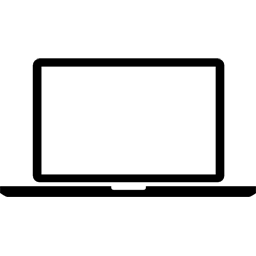}};
		\node[draw=none, line width = 1pt, above  =0.3cm of node5] (node4) {\includegraphics[width=.08\textwidth]{mix.png}};
		\node[draw=none, line width = 1pt, below  =0.3cm of node5] (node6) {\includegraphics[width=.08\textwidth]{mix.png}};
		
		\node[draw=none, line width = 1pt, left  =1.5cm of node4] (node1) {\includegraphics[width=.08\textwidth]{mix.png}};
		\node[draw=none, line width = 1pt, below =0.3cm of node1] (node2) {\includegraphics[width=.08\textwidth]{mix.png}};
		\node[draw=none, line width = 1pt, below =0.3cm of node2] (node3) {\includegraphics[width=.08\textwidth]{mix.png}};

		\node[draw=none, line width = 1pt, right  =1.5cm of node4] (node7) {\includegraphics[width=.08\textwidth]{mix.png}};
		\node[draw=none, line width = 1pt, right  =1.5cm of node5] (node8) {\includegraphics[width=.08\textwidth]{mix.png}};
		\node[draw=none, line width = 1pt, right  =1.5cm of node6] (node9) {\includegraphics[width=.08\textwidth]{mix.png}};
		
		\node[draw=none, fill=none, left =0.3cmof cloud, yshift=2cm] (provider1) {\includegraphics[width=.1\textwidth]{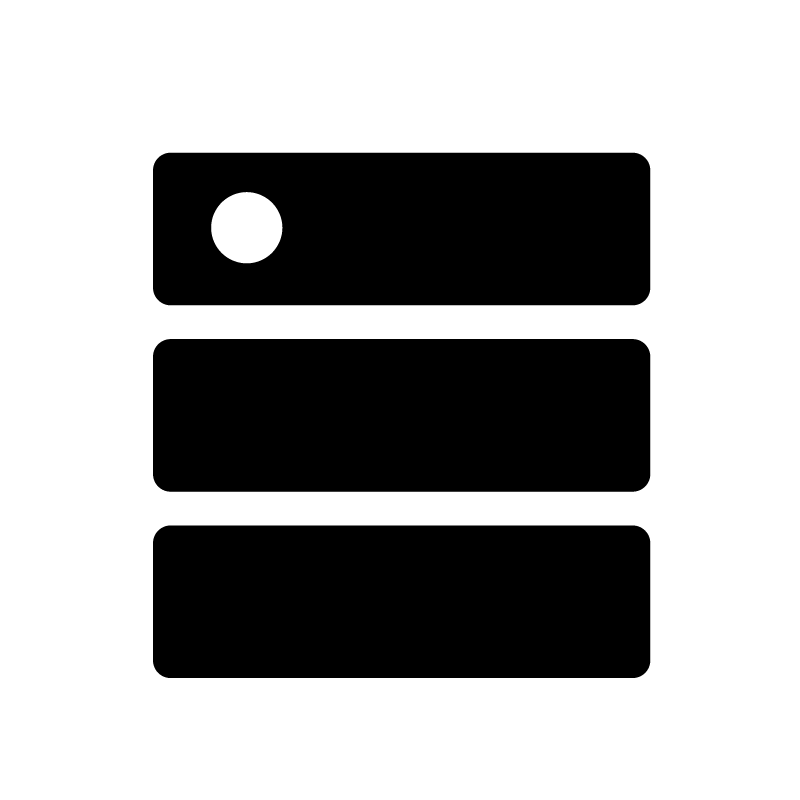}};
		\node[draw=none, fill=none, below =2cm of provider1] (provider2) {\includegraphics[width=.1\textwidth]{provider.png}};
		\node[draw=none, fill=none, right =0.3cm of cloud, yshift=2cm] (provider3) {\includegraphics[width=.1\textwidth]{provider.png}};
		\node[draw=none, fill=none, below =2cm of provider3] (provider4) {\includegraphics[width=.1\textwidth]{provider.png}};
		\node[draw=none, fill=none, left =2cm of provider1] (senderClient) {\includegraphics[width=.1\textwidth]{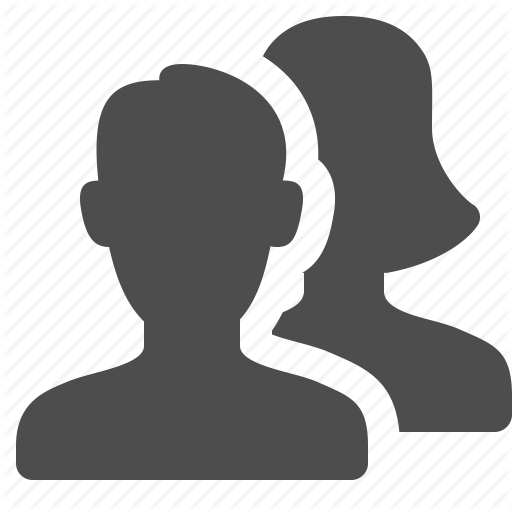}};
		\node[draw=none, fill=none, left =2cm of provider2] (senderClient2) {\includegraphics[width=.1\textwidth]{users.png}};
		\node[draw=none, fill=none, right =2cm of provider3] (receiverClient) {\includegraphics[width=.1\textwidth]{users.png}};
		\node[draw=none, fill=none, right =2cm of provider4] (receiverClient2) {\includegraphics[width=.1\textwidth]{users.png}};
		\draw[->, dashed, opacity=0.3, line width = 1pt] (node1) edge node {} (node4);
		\draw[->, dashed, opacity=0.3, line width = 1pt] (node1) edge node {} (node5);
		\draw[->, dashed, opacity=0.3, line width = 1pt](node1) edge node {} (node6);
		\draw[->, dashed, opacity=0.3, line width = 1pt] (node2) edge node {} (node4);
		\draw[->, dashed, opacity=0.3, line width = 1pt] (node2) edge node {} (node5);
		\draw[->, dashed, opacity=0.3, line width = 1pt](node2) edge node {} (node6);
		\draw[->, dashed, opacity=0.3, line width = 1pt] (node3) edge node {} (node4);
		\draw[->, dashed, opacity=0.3, line width = 1pt] (node3) edge node {} (node5);
		\draw[->, dashed, opacity=0.3, line width = 1pt](node3) edge node {} (node6);
		\draw[->, dashed, opacity=0.3, line width = 1pt] (node4) edge node {} (node7);
		\draw[->, dashed, opacity=0.3, line width = 1pt] (node4) edge node {} (node8);
		\draw[->, dashed, opacity=0.3, line width = 1pt](node4) edge node {} (node9);
		\draw[->, dashed, opacity=0.3, line width = 1pt] (node5) edge node {} (node7);
		\draw[->, dashed, opacity=0.3, line width = 1pt] (node5) edge node {} (node8);
		\draw[->, dashed, opacity=0.3, line width = 1pt](node5) edge node {} (node9);
		\draw[->, dashed, opacity=0.3, line width = 1pt] (node6) edge node {} (node7);
		\draw[->, dashed, opacity=0.3, line width = 1pt] (node6) edge node {} (node8);
		\draw[->, dashed, opacity=0.3, line width = 1pt](node6) edge node {} (node9);
		\draw[->, red, dotted, line width = 3pt] (provider1) edge node {} (node1);
		\draw[->, red, dotted,  line width = 3pt] (node1) edge node {} (node5);
		\draw[->, red, dotted,  line width = 3pt] (node5) edge node {} (node7);
		\draw[->, red, dotted, line width = 3pt, bend right] (node7) edge node {} (provider1);
		\draw[->, blue, line width = 4pt] (provider2) edge node {} (node2);
		\draw[->, blue, line width = 4pt] (node2) edge node {} (node5);
		\draw[->, blue, line width = 4pt] (node5) edge node {} (node8);
		\draw[->, blue, line width = 4pt] (node8) edge node {} (provider3);
		\draw[->, green, line width = 2pt] (node9) edge node {} (provider4);
		\draw[->, green, line width = 2pt, bend left] (provider4) edge node {} (node3);
		\draw[->, green, line width = 2pt] (node3) edge node {} (node5);
		\draw[->, green, line width = 2pt] (node5) edge node {} (node9);
		\path[->,  orange, line width = 2pt, sloped, transform canvas={yshift=1.5ex}] (senderClient) edge node[pos=0.5, above]  {\includegraphics[width=.08\textwidth]{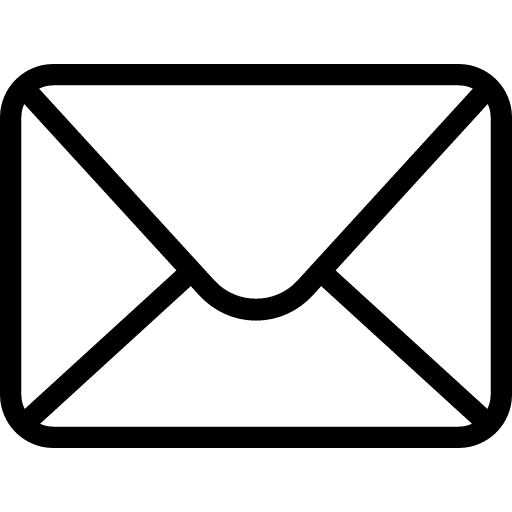}} (provider1);
		\draw[->, dashed, orange, sloped, line width = 2pt, transform canvas={yshift=-2ex}] (provider1)  [below]  edge node{\huge Storage} (senderClient);
		
		\path[->,  orange, line width = 2pt, sloped, transform canvas={yshift=1.5ex}] (senderClient2) edge node[pos=0.5, above]  {\includegraphics[width=.08\textwidth]{envelope.png}}  (provider2);
		\draw[->, dashed, orange, sloped, line width = 2pt, transform canvas={yshift=-2ex}] (provider2)  [below]  edge node{\huge Storage} (senderClient2);

		\draw[->, orange, line width = 2pt, sloped, transform canvas={yshift=1.5ex}] (receiverClient) edge node[pos=0.5, above]  {\includegraphics[width=.08\textwidth]{envelope.png}}  (provider3);
		\draw[->, dashed, orange, line width = 2pt, sloped, transform canvas={yshift=-2ex}] (provider3) [below]  edge node{\huge Storage} (receiverClient);
		
		\draw[->, orange, line width = 2pt, sloped, transform canvas={yshift=1.5ex}] (receiverClient2) edge node[pos=0.5, above]  {\includegraphics[width=.08\textwidth]{envelope.png}}  (provider4);
		\draw[->, dashed, orange, line width = 2pt, sloped, transform canvas={yshift=-2ex}] (provider4) [below]  edge node{\huge Storage} (receiverClient2);
		\node[draw=none, fill=none, above=0.5cm of cloud, align=center] (text4) {\huge Users' loop cover traffic \\ \huge generates traffic \\ \huge in two directions};
		\node[draw=none, fill=none, below right=of node9, align=left] (text1) {\huge Mixes can detect\\ \huge n-1 attacks};
		\draw[dashed, gray] (text1) edge node {} (node9);
		\node[draw=none, fill=none, below=0.6cm of provider2, align=center] (text5) {\huge Providers offer \\ \huge offline storage \\ \huge when user is offline};	
		\draw[dashed, gray] (text5) -- (provider2);	
 	\end{tikzpicture}
	}
	\centering \caption{The Loopix Architecture. Clients pass the messages to the providers, which are responsible for injecting traffic into the network. The received messages are stored in individual inboxes and retrieved by clients when they are online.}\label{fig:loopix}
\end{figure}
\end{center}

\paragraph{Path selection.}
As opposed to onion routing, in \sysname the communication path for every single message is chosen independently, even between the same pair of users. 

Messages are routed through $l$ layers of mix nodes, assembled in a stratified topology~\cite{danezis2003mix, dingledine2004synchronous}. Each mix node is connected 
only with all the mix nodes from adjacent layers. This ensures that few links are used, and those few links are well covered in traffic; stratified topologies mix well in few steps~\cite{dingledine2004synchronous}. Providers act as the first and last layer of mix servers. 
To send a message, the sender encapsulates the routing information described above into a Sphinx packet which travels through their provider, a sequence of mix servers, until it reaches the receiver's provider and finally the \emph{receiver}. For each of those hops the sender samples a delay from an exponential distribution with parameter $\mu$, and includes it in the routing information to the corresponding relay.


\paragraph{Sending messages and cover traffic.}
Users and mix servers continuously generate a bed of \emph{real} and \emph{cover traffic} that is injected into the network. 
Our design guarantees that all outgoing traffic sent by users can by modeled by a Poisson process. 

To send a message, a user packages their message into a mix packet and places it into their \emph{buffer}---a first-in-first-out (FIFO) queue that stores all the messages scheduled to be sent. 

Each sender periodically checks, following the exponential distribution with parameter $\frac{1}{\lambda_{P}}$, whether there is any scheduled message to be sent in their buffer. 
If there is a scheduled message, the sender pops this message from the buffer queue and sends it, otherwise a \emph{drop} cover message is generated (in the same manner as a regular message) and sent (depicted as the the four middle blue arrows in \Cref{fig:loopix}). Cover messages are routed through the sender's provider and a chain of mix nodes to a random destination provider. The destination provider detects the message is cover based on the special drop flag encapsulated into the packet header, and drops it. 
Thus, regardless of whether a user actually wants to send a message or not, there is always a stream of messages being sent according to a Poisson process $\pois(\lambda_{P})$. 

Moreover, independently from the above, all users emit separate streams of special indistinguishable types of \emph{cover messages}, which also follow a Poisson process.
The first type of cover messages are Poisson distributed \emph{loops} emitted at rate $\lambda_{L}$. These are routed through the network and \emph{looped back} to the senders (the upper four red arrows in \Cref{fig:loopix}), by specifying the sending user as the recipient. These \emph{``loops''} inspire the system's name. 
Users also inject a separate stream of \emph{drop} cover messages, defined before, following the Poisson distribution $\pois(\lambda_{D})$.  Additionally, each user sends at constant time a stream of \emph{pull} requests to its \emph{provider} in order to retrieve received messages, described in \Cref{sec:storing}. 

Each mix also injects their own \emph{loop} cover traffic, drawn from a Poisson process with rate $\pois(\lambda_{M})$, into the network. Mix servers inject mix packets that are looped through a path, made up of a subset of other mix servers and one randomly selected 
\emph{provider}, back to the sending mix server, creating a second type of \emph{``loop''}. This loop originates and ends in a mix server (shown as the lower four green arrows in \Cref{fig:loopix}). 
In \Cref{sec:analysis} we examine how the \emph{loops} and the \emph{drop} cover messages help protect against passive and active attacks.

\paragraph{Message storing and retrieving.}\label{sec:storing} 
Providers do not forward the incoming mix packets to users but instead buffer them. Users, when online, \emph{poll} providers or register their online status to download a fixed subset of stored messages, allowing for the reception of the off-line messages.  Recall that cover loops are generated by users and traverse through the network and come back to the sender. Cover loops serve as a cover set of \emph{outgoing} and \emph{incoming} real messages. 
Whenever a user requests messages, their provider responds with a constant number of messages, which includes their cover loop messages and real messages. If the inbox of a particular user contains fewer messages than this constant number, the provider sends dummy messages to the sender up to that number. 

\subsection{The Poisson Mix Strategy}\label{sec:poisson_mix}

\sysname leverages cover traffic to resist traffic analysis while still achieving low- to mid-latency. To this end \sysname employs a mixing strategy that we call a \emph{Poisson Mix}, to foil observers from learning about the correspondences between input and output messages. The Poisson Mix is a simplification of the Stop-and-go mix strategy~\cite{kesdogan1998stop}. A similar strategy has been used to model traffic in onion routing servers~\cite{DanezisPet04}. In contrast, recall that in \sysname each message is source routed through an independent route in the network.

The Poisson Mix functions as follows: mix servers listen for the incoming mix packets and received messages are checked for duplication and decoded using the mix node's private keys. The detected duplicates are dropped. Next, the mix node extracts a subsequent mix packet. Decoded mix packets are not forwarded immediately, but each of them is delayed according to a source pre-determined delay $d_{i}$. Honest clients chose those delays, independently for each hop, from an exponential distribution with a parameter $\mu$. We assume that this parameter is public and the same for all mix nodes. 



%
\begin{center}
\begin{figure}[t!]
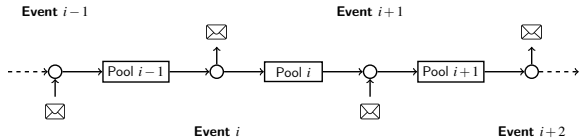

	\resizebox{\columnwidth}{!}{
	\begin{tikzpicture}[node distance=1cm,font=\sffamily,
block/.style={
draw,
rectangle, 
minimum width={width("Pool $i+1$")+2pt},
line width = 1pt}]

	\node[draw=none, fill=none] (start) {};

		\node[draw, circle, line width = 1pt, right = 1 cm of start] (c1) {};
	\node[draw=none, fill=none, below of=c1] (m1) {\includegraphics[width=.03\textwidth]{envelope.png}};

		\node[draw=none, fill=none, node distance=1.5cm, above of=c1] (l1) {\textsf{\textbf{Event $i-1$}}};

		\node[block, right = 1cm of c1] (poolim1) {Pool $i-1$};
		\node[draw, circle, line width = 1pt,right = 1cm of poolim1] (c2) {};
		\node[draw=none, fill=none, node distance=1.5cm, below of=c2] (l2) {\textsf{\textbf{Event $i$}}};

	\node[draw=none, fill=none, above of=c2] (m2) {\includegraphics[width=.03\textwidth]{envelope.png}};
		\node[block, right = 1cm of c2] (pooli) {Pool $i$};
		\node[draw, circle, line width = 1pt, right = 1cm of pooli] (c3) {};
	\node[draw=none, fill=none, below of=c3] (m3) {\includegraphics[width=.03\textwidth]{envelope.png}};
		\node[draw=none, fill=none, node distance=1.5cm, above of=c3] (l2) {\textsf{\textbf{Event $i+1$}}};

		\node[block, right = 1 cm of c3] (poolip1) {Pool $i+1$};
		\node[draw, circle, line width = 1pt, right = 1 cm of poolip1] (c4) {};
	\node[draw=none, fill=none, above of=c4] (m4) {\includegraphics[width=.03\textwidth]{envelope.png}};
			\node[draw=none, fill=none, node distance=1.5cm, below of=c4] (l2) {\textsf{\textbf{Event $i+2$}}};

	\node[draw=none, fill=none, right = 1cm of c4] (end) {};

	\draw[->, black, dashed, line width = 1pt]  (start) -- (c1);
	\draw[->, black, line width = 1pt]  (m1) -- (c1);
	\draw[->, black, line width = 1pt]  (c1) -- (poolim1);
	\draw[->, black, line width = 1pt]  (poolim1) -- (c2);
	\draw[->, black, line width = 1pt]  (c2) -- (m2);
	\draw[->, black, line width = 1pt]  (c2) -- (pooli);
	\draw[->, black, line width = 1pt]  (pooli) -- (c3);
	\draw[->, black, line width = 1pt]  (m3) -- (c3);
	\draw[->, black, line width = 1pt]  (c3) -- (poolip1);
	\draw[->, black, line width = 1pt]  (poolip1) -- (c4);
	\draw[->, black, line width = 1pt]  (c4) -- (m4);
	\draw[->, black, dashed, line width = 1pt]  (c4) -- (end);
 	\end{tikzpicture}
	}
	\centering \caption{The Poisson Mix strategy mapped to a Pool mix strategy. Each single message sending or receiving event leads to a new pool of messages that are exchangeable and indistinguishable with respect to their departure times. } \label{fig:poisson}
\end{figure}
\end{center}

\vspace{-8mm}
\paragraph{Mathematical model of a Poisson Mix.}\label{sec:poisson_mix}
Honest clients and mixes generate drop cover traffic, loop traffic, and messaging traffic following a Poisson process. Aggregating Poisson processes results in a Poisson process with the sum of their rates, therefore we may model the streams of traffic received by a Poisson mix as a Poisson process. It is the superposition of traffic streams from multiple clients. It has a rate $\lambda_n$ depending on the number of clients and the number of mix nodes. 

Since this input process is a Poisson process and each message is independently delayed using an exponential distribution with parameter $\mu$, the Poisson Mix may be modeled as an $M/M/\infty$ queuing system -- for we have a number of well known theorems~\cite{bolch2006queueing}.
%
We know that output stream of messages is also a Poisson process with the parameter $\lambda_n$ as the the input process. 
We can also derive the distribution of the number of messages within the Poisson Mix~\cite{mitzenmacher2005probability}:
\begin{lemma}\label{lemma_avg}
	The mean number of messages in the Poisson Mix with input Poisson process $\pois(\lambda)$ and exponential delay parameter $\mu$ at a steady state follows the Poisson distribution $\pois\left(\lambda/\mu\right)$.
\end{lemma}
Those characteristics give the Poisson Mix its name. This allows us to calculate the mean number of messages \emph{perfectly} mixed together at any time, as well as the probability that the number of messages falls below or above certain thresholds. 

The Poisson Mix, under the assumption that it approximates an $M/M/\infty$ queue is a stochastic variant of a pool mixing strategy~\cite{serjantov2003anonymity}. Conceptually, each message sending or receiving leads to a pool within which messages are indistinguishable due to the memoryless property of the exponential delay distribution.
\begin{lemma}[Memoryless property~\cite{mitzenmacher2005probability}]
For an exponential random variable $X$ with parameter $\mu$ holds
$\Pr[X > s + t | X > t] = \Pr[X > s].$
\end{lemma}
Intuitively, any two messages in the same pool are emitted next with equal probability -- no matter how long they have been waiting. As illustrated in \Cref{fig:poisson}, the receiving event $i-1$ leads to a pool of messages $i-1$, until the sending event $i$. From the perspective of the adversary observing all inputs and outputs, all messages in the pool $i-1$ are indistinguishable from each other. Only the presence of those messages in the pool is necessary to characterize the hidden state of the mix (not their delay so far). 
Relating the Poisson mix to a pool mix allows us to compute easily and exactly both the entropy metric for the anonymity it provides~\cite{serjantov2002towards} within a trace (used in \Cref{sec:poisson_sec}). It also allows us to compute the likelihood that an emitted message was any specific input message used in our security evaluation.

\section{Analysis of \sysname security properties}\label{sec:analysis} 

In this section we present the analytical and experimental evaluation of the security of \sysname and argue its resistance to traffic analysis and active attacks. 

\subsection{Passive attack resistance}\label{sec:system_analysis}

\subsubsection{Message Indistinguishability}\label{sec:message_indistinguishability}
\sysname relies on the Sphinx packet format~\cite{danezis2009sphinx} to provide bitwise unlinkability of incoming and outgoing messages from a mix server; it does not leak information about the number of hops a single message has traversed or the total path length; and it is resistant to tagging attacks.

For \sysname, we make minor modifications to Sphinx
to allow auxiliary meta-information to be passed to different mix servers. 
Since all the auxiliary information is encapsulated into the header of the packet in the same manner 
as any meta-information was encapsulated in the Sphinx design, the security properties are unchanged. 
An external adversary and a corrupt intermediate mix node or a corrupt provider will not be able to distinguish \emph{real} messages from \emph{cover} messages of any type. 
Thus, the GPA observing the network cannot infer any information about the type of the transmitted messages, and intermediate nodes cannot distinguish real messages, drop cover messages or loops of clients and other nodes from each other.
Providers are able to distinguish \emph{drop} cover message destined for them from other messages, since they learn the \emph{drop flag} attached in the header of the packet. 
Each mix node learns the delay chosen by clients for this particular mix node, but all delays are chosen independently from each other.

\subsubsection{Client-Provider unobservability}\label{sec:client-provider-unobservability}

In this section, we argue the \emph{sender and receiver unobservability} against different adversaries in our threat model. Users emit payload messages following a Poisson distribution with parameter $\lambda_{P}$. All messages scheduled for sending by the user are placed within a first-in-first-out buffer.
According to a Poisson process, a single message is popped out of the buffer and sent, or a drop cover message is sent in case the buffer is empty. 
Thus, from an adversarial perspective, there is always traffic emitted modeled by $\pois\left(\lambda_{P}\right)$. Since clients send also streams of cover traffic messages with rates $\lambda_L$ for loops and $\lambda_D$ for drop cover messages, the traffic sent by the client follows $\pois\left(\lambda_{P} + \lambda_{L} + \lambda_{D}\right)$. Thus we achieve perfect \emph{sender unobservability}, since the adversary cannot tell whether a genuine message or a drop cover message is sent.

\begin{center}
\begin{figure}[t!]
	\resizebox{\columnwidth}{!}{
	\begin{tikzpicture}[font=\sffamily]
		\node[draw=none, fill=none, yshift=0.6cm] (inbox1) {\scriptsize Inbox I \hspace{4mm} \includegraphics[width=.02\textwidth]{envelope.png} \includegraphics[width=.02\textwidth]{envelope.png} \includegraphics[width=.02\textwidth]{envelope.png} };
		\node[draw=none, fill=none, below= 0.005cm of inbox1] (inbox2) {\scriptsize Inbox II \hspace{12mm} \includegraphics[width=.02\textwidth]{envelope.png} };
		\node[draw=none, fill=none, below= 0.005cm of inbox2] (inbox3) {\scriptsize Inbox III \hspace{7mm} \includegraphics[width=.02\textwidth]{envelope.png} \includegraphics[width=.02\textwidth]{envelope.png} };
		\node[draw=none, fill=none, right = 1.6cm of inbox2] (provider2) {\includegraphics[width=.08\textwidth]{provider.png}};
		\node[draw=none, fill=none, right= 1cm of provider2] (user2) {\includegraphics[width=.02\textwidth]{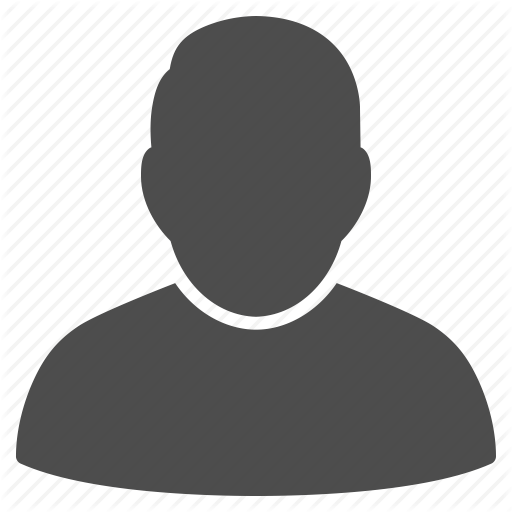}};
		\node[draw=none, fill=none, above= 0.5cm of user2] (user1) {\includegraphics[width=.02\textwidth]{single_user.png}};
		\node[draw=none, fill=none, below= 0.5cm of user2] (user3) {\includegraphics[width=.02\textwidth]{single_user.png}};

		\draw[blue,thick, dashed, rounded corners,  line width = 1.0pt, opacity=0.5] ($(inbox1.north west)+(-0.3,-0)$)  rectangle ($(provider2.south west)+(0.5,-0.1)$);
		
		\draw [->, line width=1.0pt, right= 1cm of provider2, sloped] (provider2.north east) -- (user1) node[pos=0.5, below] {\includegraphics[width=.015\textwidth]{envelope.png} \includegraphics[width=.015\textwidth]{envelope.png} \includegraphics[width=.015\textwidth]{envelope.png}};
		\draw [->, line width=1.0pt, sloped] (provider2) -- (user2) node[pos=0.5, below] {\includegraphics[width=.015\textwidth]{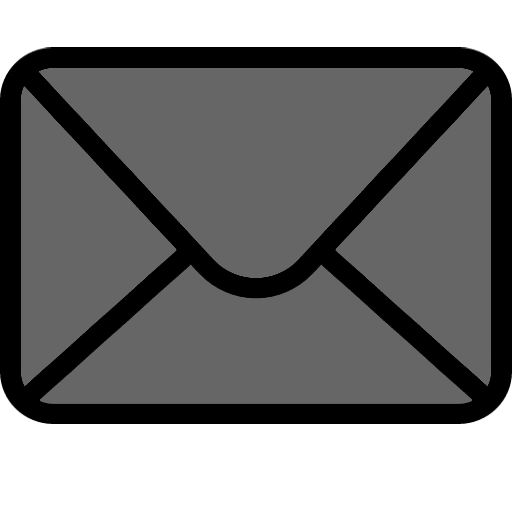} \includegraphics[width=.015\textwidth]{envelope.png} \includegraphics[width=.015\textwidth]{envelope_fake.png}};
		\draw [->, line width=1.0pt, sloped] (provider2.south east) -- (user3)  node[pos=0.5, below] {\includegraphics[width=.015\textwidth]{envelope_fake.png} \includegraphics[width=.015\textwidth]{envelope.png} \includegraphics[width=.015\textwidth]{envelope.png}};
		\node[draw=none, fill=none, above = 0.5 cm of provider2, xshift=1.2cm] (adversary) {\includegraphics[width=.03\textwidth]{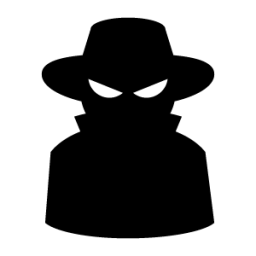}};		
		
	\end{tikzpicture}
	}
	\centering \caption{Provider stores messages destined for assigned clients in a particular inbox. When users pull messages from the mix node, the provider generates cover messages to guarantee that the adversary cannot learn how many messages are in the users inbox. The messages from the inbox and dummies are indistinguishable.}
	\label{fig:retrieving_messages}
\end{figure}
\end{center}

When clients query providers for received messages, the providers always send a constant number of messages to the client. 
If the number of messages in client's inbox is smaller than a constant threshold, the provider generates additional dummy messages.
Thus, the adversary observing the client-provider connection, as presented on \Cref{fig:retrieving_messages}, cannot learn how many messages were in the user's inbox. Note that, as long as the providers are honest, the protection and \emph{receiver unobservability} is perfect and the adversary cannot learn any information about the inbox and outbox of any client. 

If the provider is dishonest, then they are still uncertain whether a received message is genuine or the result of a client loop -- something that cannot be determined from their bit pattern alone. However, further statistical attacks may be possible, and we leave quantifying exact security against this threat model as future work.

\subsubsection{Poisson mix security}\label{sec:poisson_sec}

We first show that a single honest Poisson mix provides a measure of \emph{sender-receiver unlinkability}.
From the properties of Poisson mix, we know that the number of messages in the mix server at a steady state depends on the ratio of the incoming traffic ($\lambda$) and the delay parameter $(\mu)$ (from \Cref{sec:poisson_mix}). 
The number of messages in each mix node at any time will on average be $\frac{\lambda}{\mu}$. However, an adversary observing the messages flowing into and out of a single mix node could estimate the exact number of messages within a mix with better accuracy -- hindered only by the mix loop cover traffic.


We first consider, conservatively, the case where a mix node is not generating any loops and the adversary can count the exact number of messages in the mix. Let us define $o_{n,k,l}$ as an adversary $A$ observing a mix in which $n$ messages arrive and are mixed together. The adversary then observes an outgoing set of $n-k$ messages and can infer that there are now $k < n$ messages in the mix. 
Next, $l$ additional messages arrive at the mix before any message leaves, and the pool now mixes $k+l$ messages. 
The adversary then observes exactly one outgoing message $m$ and tries to correlate it with any of the $n+l$ messages which she has observed arriving at the mix node.

The following lemma is based on the memoryless property of the Poisson mix. It provides an upper bound on the probability that the adversary $A$ correctly links the outgoing message $m$ with one of the previously observed arrivals in observation $o_{n,k,l}$.
\begin{theorem}\label{probability1} 
Let $m_{1}$ be any of the initial $n$ messages in the mix node in scenario $o_{n,k,l}$, and let $m_{2}$ be any of the $l$ messages that arrive later. Then 
\begin{align}
Pr(m = m_{1}) &= \frac{k}{n(l+k)}, \\
Pr(m = m_{2}) &= \frac{1}{l+k}.
\end{align}
\end{theorem}

Note that the last $l$ messages that arrived at the mix node have equal probabilities of being the outgoing message $m$, independently of their arrival times. Thus, the arrival and departure times of the messages cannot be correlated, and the adversary learns no additional information by observing the timings. Note that $\frac{1}{l+k}$ is an upper bound on the probability that the adversary $A$ correctly links the outgoing message to an incoming message. Thus, continuous observation of a Poisson mix leaks no additional information other than the number messages present in the mix.
%
We leverage those results from about a single Poisson Mix to simulate the information propagated withing a the whole network observed by the adversary (c.f.~\Cref{sec:secsim}).

We quantify the anonymity of messages in the mix node empirically, using an information theory based metric introduced in~\cite{serjantov2002towards, diaz2002towards}. 
We record the traffic flow for a single mix node and compute the distribution of probabilities that the outgoing message is the adversary's target message. Given this distribution we compute the value of Shannon entropy (see~\Cref{sec:appendix}), a measure of unlinkability of incoming to outgoing messages. We compute this using the $\mathsf{simpy}$ package in $\mathsf{Python}$. All data points are averaged over $50$ simulations.

\begin{figure}[t!]
\small
\centering
  \includegraphics[width=\columnwidth]{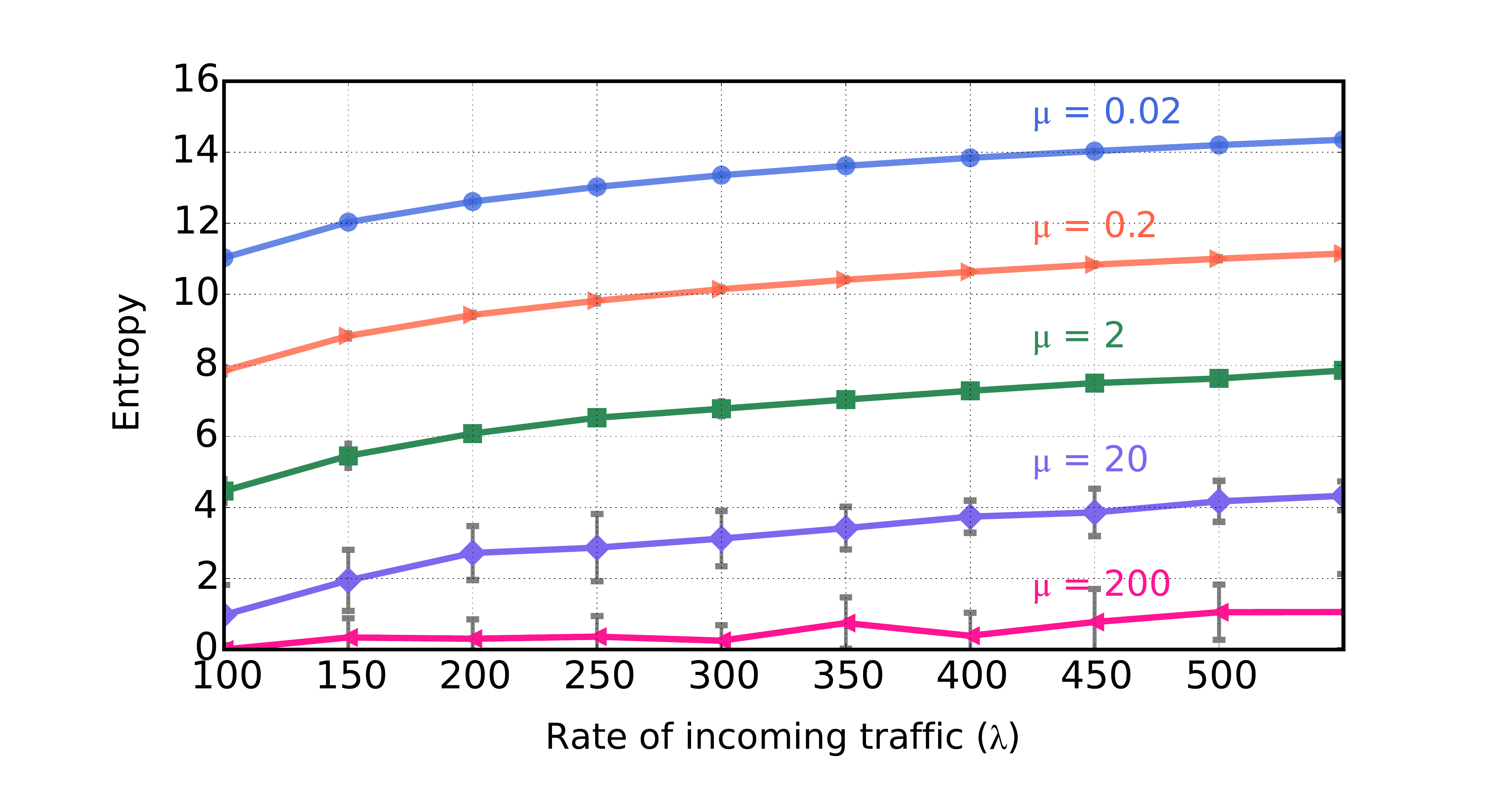}
  \caption{Entropy versus the changing rate of the incoming traffic for different delays with mean $\frac{1}{\mu}$. In order to measure the entropy we run a simulation of traffic arriving at a single \sysname mix node.}
  \label{fig:entropy}
\end{figure}

\Cref{fig:entropy} depicts the change of entropy against an increasing rate of incoming mix traffic $\lambda$. 
We simulate the dependency between entropy and traffic rate for different mix delay parameter $\mu$ by recording the traffic flow and changing state of the mix node's pool. As expected, we observe that for a fixed delay, 
the entropy increases when the rate of traffic increases. Higher delay also results in an increase in entropy, denoting a larger potential anonymity set, since more messages are mixed together.

In case the mix node emits loop cover traffic, the adversary with observation $o_{n,k,l}$, tries to estimate the probability that the observed outgoing message is a particular \emph{target} message she observed coming into the mix node. An outgoing message can be either input message or a loop message generated by the mix node -- resulting in additional uncertainty for the adversary.

\begin{theorem}\label{probability2} 
Let $m_{1}$ be any of the initial $n$ messages in the mix node in scenario $o_{n,k,l}$, and let $m_{2}$ be any of the $l$ messages that arrive later. 
Let $\lambda_{M}$ denote the rate at which mix node generates loop cover traffic. Then, 
\begin{align*}
	\Pr(m = m_{2}) &= \frac{k}{n}\cdot \frac{\mu}{(l+k)\mu + \lambda_{M}}, \\
	\Pr(m = m_{1}) &= \frac{\mu}{(l+k)\mu + \lambda_{M}}.
\end{align*}
\end{theorem} 
We refer to \Cref{sec:appendix} for the proof. 
We conclude that the loops generated by the mix node obfuscate the adversary's view and decrease the probability of successfully linking input and output of the mix node. In \Cref{sec:traffic_analysis} we show that those types of loops also protect against active attacks.


\subsection{Active-attack Resistance}\label{sec:traffic_analysis}

\Cref{lemma_avg} gives the direct relationship between the expected number of messages in a mix node, the
rate of incoming traffic, and the delay induced on a message while transiting through a mix. By increasing the rate of cover traffic,
$\lambda_D$ and $\lambda_L$, users can collectively maintain strong anonymity with low message delay. 
However, once the volume of real communication traffic $\lambda_P$ increases, users can tune down the rate of cover traffic in comparison to the real traffic, while maintaining a small delay and be confident their messages are mixed with a sufficient number of messages.

In the previous section, we analyze the security properties of \sysname when the adversary observes the state of a single mix node and the traffic flowing through it. We showed, that the adversary's advantage is bounded due to the indistinguishability of messages and the memoryless property of the Poisson mixing strategy. We now investigate how \sysname can protect users' communications against active adversaries conducting the $(n-1)$ attack.


\subsubsection{Active attacks}
We consider an attack at a mix node where an adversary blocks all but a target message from entering in order
to follow the target message when it exits the mix node. This is referred to as an \emph{(n-1) attack}~\cite{serjantov2002trickle}.

A mix node needs to distinguish 
between an active attack and loop messages dropped due to congestion. We assume that each mix node chooses some public parameter $r$, which is a fraction of the number of loops that are expected to return.
If the mix node does not see this fraction of loops returning they alter
their behavior. In extremis such a mix could refuse to emit any messages -- but this would escalate this attack to full denial-of-service. A gentler approach involves generating more cover traffic on outgoing links~\cite{danezis2003heartbeat}.

To attempt an \emph{(n-1) attack}, the adversary could simply block all incoming messages to the mix node except for a target message. The \sysname mix node can notice that the self-loops are not returning and deduce it is under attack. Therefore, an adversary that wants 
to perform a stealthy attack has to be judicious when blocking messages, to ensure that a fraction $r$ of loops return to the mix 
node, i.e.\ the adversary must distinguish loop cover traffic
from other types of traffic. 
However, traffic generated by mix loops is indistinguishable from other network traffic and they cannot do this better than by chance.
Therefore given a threshold $r = \frac{\lambda_{M}}{s}, s\in\mathbb R_{>1}$ of
expected returning loops when a mix observes fewer returning it deploys appropriate countermeasures.

We analyze this strategy: since the adversary cannot distinguish loops from other traffic the adversary can 
do no better than block traffic uniformly such that a fraction $R=\frac{\lambda}{s}=\frac{\lambda_R + \lambda_M}{s}$ enter the mix, where
$\lambda_R$ is the rate of incoming traffic that is not the mix node's loops.
If we assume a steady state, the target message can expect to be mixed with $\frac{\lambda_R}{s \cdot \mu}$ messages that entered this mix, and 
$\frac{\lambda_M}{\mu}$ loop messages generated at the mix node.
Thus, the probability of correctly blocking a sufficient number of messages entering the mix node so as not to alter the behavior of the mix is:
\begin{align*}
\Pr(x = \text{target}) = \frac{1}{\lambda_R / s \cdot \mu + \lambda_M / \mu} = \frac{s \mu}{s \lambda_M + \lambda_R} 
\end{align*}

Due to the stratified topology, providers are able to distinguish mix loop messages sent from other traffic, since they are unique in not being routed to or from a client. This is not a substantial attack vector since  
mix loop messages are evenly distributed among all providers, of which a small fraction are corrupt and providers do not learn which mix node sent the loop to target it. 





\subsection{End-to-End Anonymity Evaluation}\label{sec:secsim}

We evaluate the \emph{sender-receiver third-party unlinkability} of the full \sysname system through 
an empirical analysis of the propagation of messages in the network. 
Our key metric is the \emph{expected difference in likelihood} that a message leaving 
the last mix node is sent from one sender in comparison to another sender. Given two 
probabilities $p_0 = \Pr[S_0]$ and $p_1 = \Pr[S_1]$ that the message was sent by senders $S_0$ and $S_1$, respectively, we calculate 
\begin{equation}\label{eq:simulationeps}
\varepsilon = \left| \log\left( p_0 / p_1 \right) \right|.
\end{equation}

To approximate the probabilities $p_0$ and $p_1$, we proceed as follows. 
We simulate $U = 100$ senders that generate and send messages 
(both payload and cover messages) with a rate $\lambda = 2$. 
Among them are two challenge senders $S_0$ and $S_1$ that send payload messages at 
a constant rate, i.e, they add one messages to their sending buffer every time unit.

Whenever a challenge sender $S_0$ or $S_1$ sends a payload message from its buffer, 
we tag the message with a label $S_0$ or $S_1$, respectively. All other messages, including 
messages from the remaining $98$ clients and the cover messages of $S_0$ and $S_1$ are unlabeled. 
At every mix we track the probability that an outgoing message is labeled $S_0$ or $S_1$, 
depending on the messages that entered the mix node and the number of messages that already left 
the mix node, as in \Cref{probability1}. Thus, messages leaving a mix node 
carry a probability distribution over labels $S_0$, $S_1$, or `unlabeled'. Corrupt mix nodes, assign to outgoing messages their input distributions. The probabilities naturally 
add up to $1$. For example, a message leaving a mix can be labeled as $\{S_0: 12\%, S_1: 15\%, \mathrm{unlabeled}: 73\%\}$. 

In a burn-in phase of 2500 time units, the 98 senders without $S_0$ or $S_1$ communicate. Then we start the two challenge senders and then simulate the network 
for another 100 time units, before we compute the \emph{expected difference in likelihood} metric.
We pick a final mix node and using probabilities of labels $S_0$ and $S_1$ for any message in the pool we calculate $\varepsilon$ as in \Cref{eq:simulationeps}.

This is a conservative approximation: we tell the adversary which of the messages leaving 
senders $S_0$ and $S_1$ are payload messages; and we do not consider 
mix or client loop messages confusing them.~\footnote{The soundness of our simplification 
can be seen by the fact that we could tell the adversary which messages are loops and the 
adversary could thus ignore them. This is equivalent to removing them, as an adversary could also simulate 
loop messages.} However, when we calculate our anonymity metric at a mix node we assume this mix node to be honest.

\subsubsection{Results}
We compare our metric for different parameters: depending on the delay parameter $\mu$, the number of layers in our topology $l$ and the percentage of corrupted mix nodes in the network. 
All of the below simulations are averaged over 100 repetitions and the error bars are defined by the standard deviation.

\begin{figure}[t!]
\centering
  \includegraphics[width=\columnwidth]{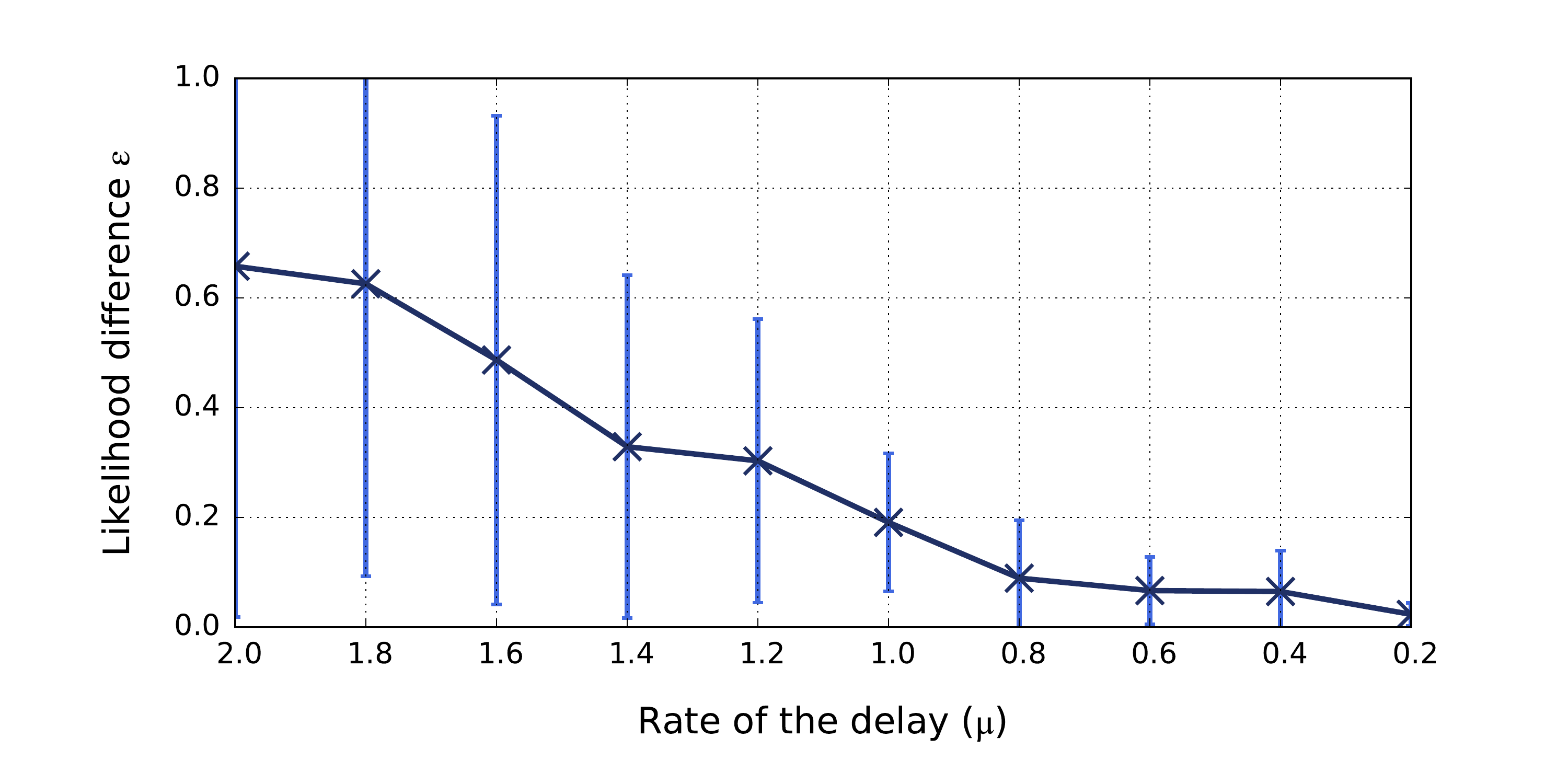}
  \caption{Likelihood difference $\varepsilon$ depending on the delay parameter $\mu$ of mix nodes. We use $\lambda=2$,  a topology of 3 layers with 3 nodes per layer and no corruption.}
  \label{fig:simgraph-mu}
\end{figure}

\paragraph{Delay.} Increasing the average delay (by decreasing parameter $\mu$) with respect to the rate of message sending $\lambda$ immediately increases anonymity (decreases $\varepsilon$) (\Cref{fig:simgraph-mu}). For $\mu = 2.0$ and $\lambda / \mu = 1$, \sysname still provides a weak form of anonymity. As this fraction increases, the log likelihood ratio grow closer and closer to zero. We consider values $\lambda / \mu \geq 2$ to be a good choice in terms of anonymity.

\begin{figure}[t!]
\centering
  \includegraphics[width=\columnwidth]{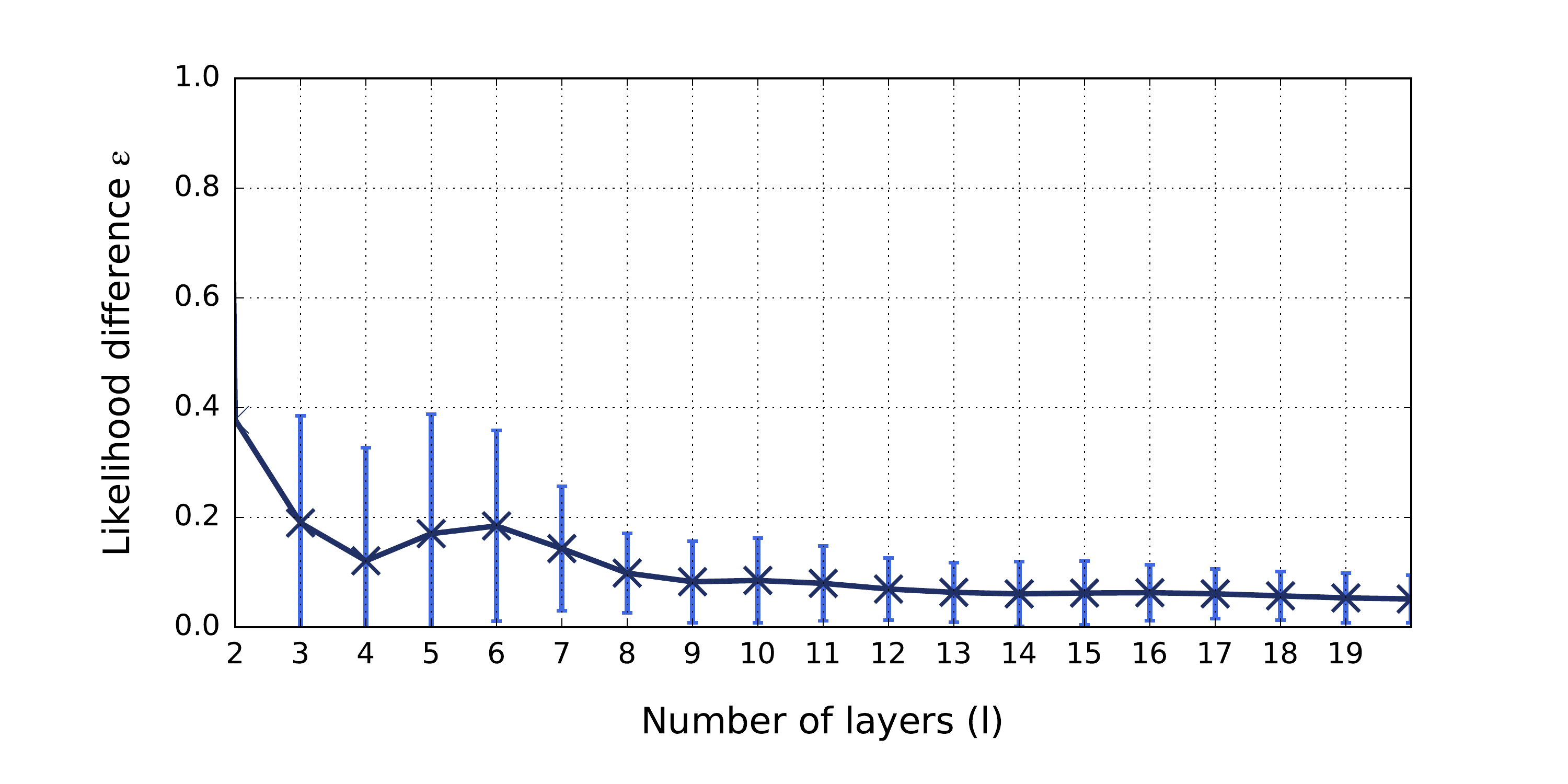}
  \caption{Likelihood difference $\varepsilon$ depending on the number of layers of mix nodes with 3 mix nodes per layer. We use $\lambda=2$, $\mu=1$, and no corruption.}
  \label{fig:simgraph-layers}
\end{figure}

\paragraph{Number of layers.} 
By increasing the number of layers of mix nodes, we can further strengthen the anonymity of \sysname users. As expected, using only one or two layers of mix nodes leads to high values of adversary advantage $\varepsilon$. An increasing number of layers $\varepsilon$ approaches zero (\Cref{fig:simgraph-layers}). We consider a number of $3$ or more layers to be a good choice. We believe the bump between 5--8 layers is due to messages not reaching latter layers within 100 time units, however results from experiments with increased duration do not display such a bump.

\begin{figure}[t!]
\centering
  \includegraphics[width=\columnwidth]{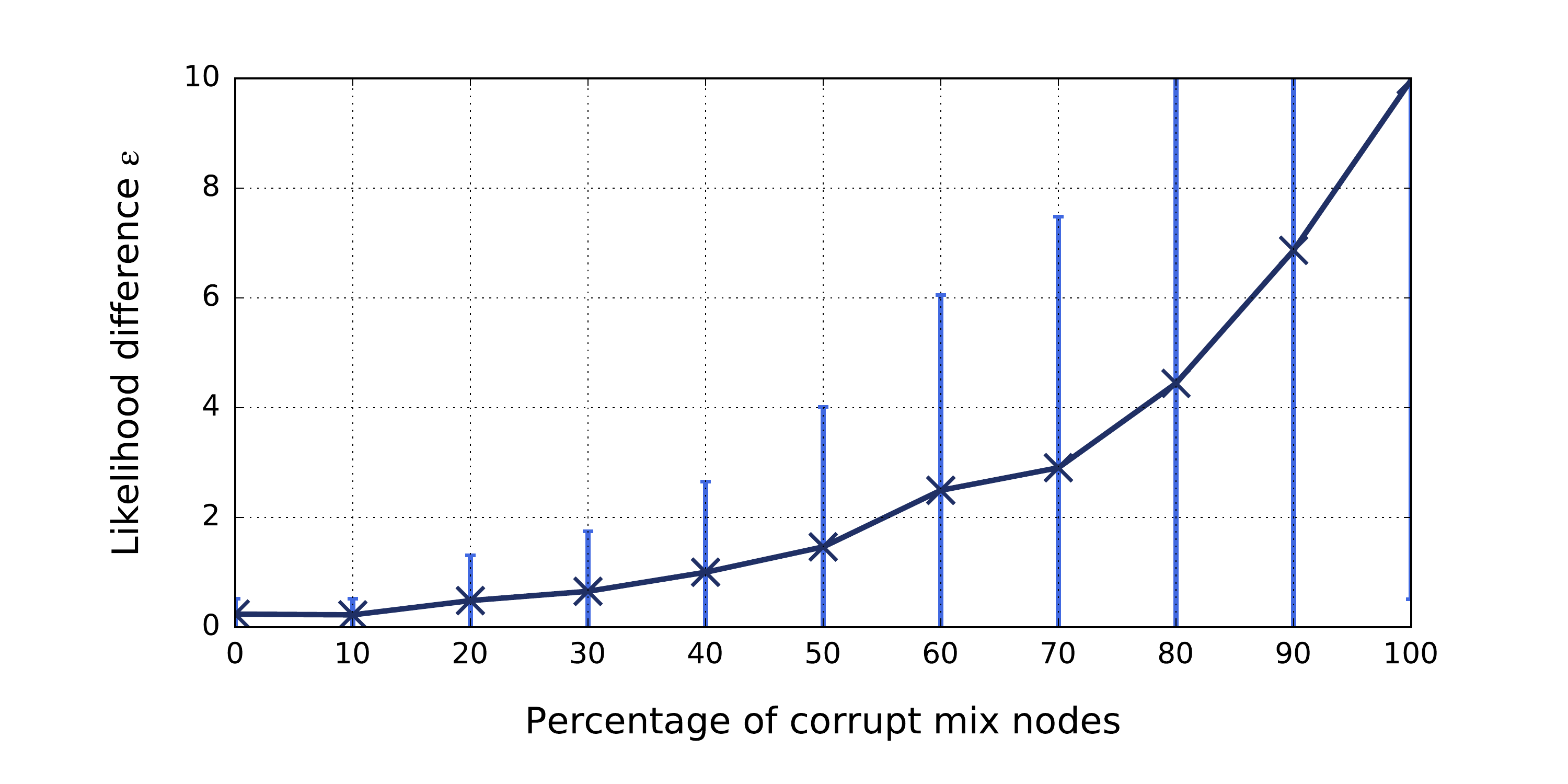}
  \caption{Likelihood difference $\varepsilon$ depending on the percentage of (passively) corrupted mix nodes. We use $\lambda=2$, $\mu=1$ and a topology of 3 layers with 3 nodes per layer.}
  \label{fig:simgraph-corruption}
\end{figure}

\paragraph{Corruption.} Finally, we analyze the impact that corrupt mix nodes have on the adversary advantage $\varepsilon$ (\Cref{fig:simgraph-corruption}). We assume that the adversary randomly corrupts mix nodes. Naturally, the advantage $\varepsilon$ increases with the percentage of corrupt mix nodes in the network. In a real-world deployment we do not expect a large fraction of mix nodes to be corrupt. While the adversary may be able to observe the entire network, to control a large number of nodes would be more costly.


\section{Performance Evaluation}\label{sec:performance}

\paragraph{Implementation.}\label{sec:implementation}

We implement the \sysname system prototype in 4000 lines of \textsf{Python 2.7} code for \emph{mix nodes}, \emph{providers} and \emph{clients}, including unit-tests, deployment, and orchestration code. \sysname source code is available under an open-source license\footnote{Public Github repository URL obscured for review.}. 
We use the \textsf{Twisted 15.5.0} network library for networking; as well as the \textsf{Sphinx} mix packet format\footnote{http://sphinxmix.readthedocs.io/en/latest/} and the cryptographic tools from the \emph{petlib}\footnote{http://petlib.readthedocs.org} library. We modify \textsf{Sphinx} to use \textsf{NIST/SEGS-p224} curves 
and to accommodate additional information inside the packet, including the delay for each hop and auxiliary flags. We also optimize its implementation leading to processing times per packet of less than $1ms$.
 
The most computationally expensive part of \sysname is messages processing and packaging, which involves cryptographic operations. Thus, we implement \sysname as a multi-thread system, with cryptographic processing happening in a thread pool separated from the rest of the operations in the main thread loop. To recover from congestion we implement active queue management based on a PID controller and we drop messages when the size of the queue reaches a (high) threshold.


\paragraph{Experimental Setup.}
We present an experimental performance evaluation of the \sysname system running on the {\tt AWS EC2} platform. All mix nodes and providers run as separate instances. Mix nodes are deployed on {\tt m4.4xlarge} instances running {\tt EC2 Linux} on $2.3\,GHz$ machines with $64\,GB$ RAM memory. Providers, since they handle more traffic, storage and operations, are deployed on {\tt m4.16xlarge} instances with $256\,GB$ RAM. We select large instances to ensure that the providers are not the bottleneck of the bandwidth transfer, even when users send messages at a high rate. This reflects real-world deployments where providers are expected to be well-resourced. We also run one {\tt m4.16xlarge} instance supporting 500 clients. We highlight that each client runs as a separate process and uses a unique port for transporting packets. Thus, our performance measurements are obtained by simulating a running system with independent clients\footnote{Other works, e.g.~\cite{tyagi2016stadium, van2015vuvuzela}, report impressive evaluations in terms of scale, but in fact are simple extrapolations and not based on empirical results.
}. In order to measure the system performance, we run six mix nodes, arranged in a stratified topology with three layers, each layer composed of two mix nodes. Additionally, we run four providers, each serving approximately $125$ clients.
The delays of all the messages are drawn from an exponential distribution with parameter $\mu$, which is the same for all mix servers in the network.
All measurements are take from network traffic dumps using {\tt tcpdump}.  

\begin{figure}[t!]
\centering
\includegraphics[width=\columnwidth]{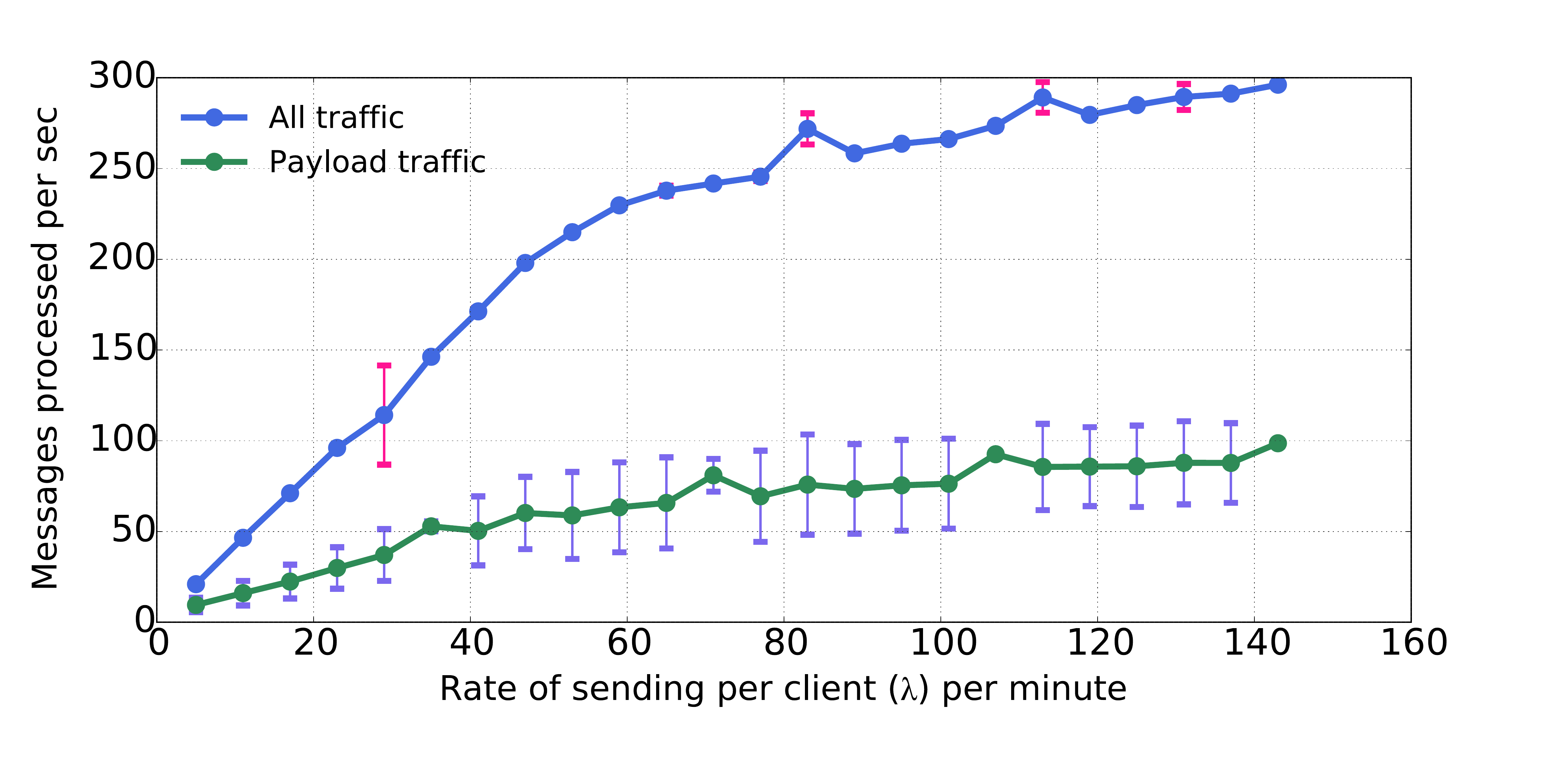}
\caption{Overall bandwidth and good throughput per second for a single mix node.}
\label{fig:mixbandwidth}
\end{figure}

\paragraph{Bandwidth.}
First, we evaluate the maximum bandwidth of mix nodes by measuring the rate at which a single mix node processes messages, for an increasing overall rate at which users send messages.

We set up the fixed delay parameter $\mu = 1000$ (s.t. the average delay is $1\,ms$). We have $500$ clients actively sending messages at rate $\lambda$ each, which is the sum of payload, loop and drop rates, i.e., $\lambda = \pois(\lambda_{L} + \lambda_{D} + \lambda_{P})$. 
We start our simulation with parameters $\lambda_{L} = \lambda_{D} = 1$ and $\lambda_{P} = 3$ messages per minute for a single client. Mix nodes send loop cover traffic at rate starting from $\lambda_{M} = 1$. Next, we periodically increase each Poisson rate by another $2$ messages per minute. 

In order to measure the overall bandwidth, i.e.\ the number of all messages processed by a single mix node, we use the network packet analyzer {\tt tcpdump}. Since real and cover message packets are indistinguishable, we measure the good throughput by encapsulating an additional, temporary, \emph{typeFlag} in the packet header for this evaluation, which leaks to the mix the message type---real or cover---and is recorded. 
 
\Cref{fig:mixbandwidth} illustrates the number of total messages and the number of payload messages that are processed by a single mix node versus the overall sending rate $\lambda$ of a single user.  
We observe that the bandwidth of the mix node increases linearly until it reaches around $225$ messages per second. After that point the performance of the mix node stabilizes and we observe a much smaller growth.  
We highlight that the amount of \emph{real} traffic in the network depends on the parameter $\lambda_{P}$ within $\lambda$. A client may chose to tune up the rate of real messages sent, by tuning down the rate of loops and drop messages -- at the potential loss of security in case less cover traffic is present in the system overall. Thus, depending on the size of the honest user population in \sysname, we can increase the rate of goodput.

\begin{figure}[t!]
\centering
\includegraphics[width=\columnwidth]{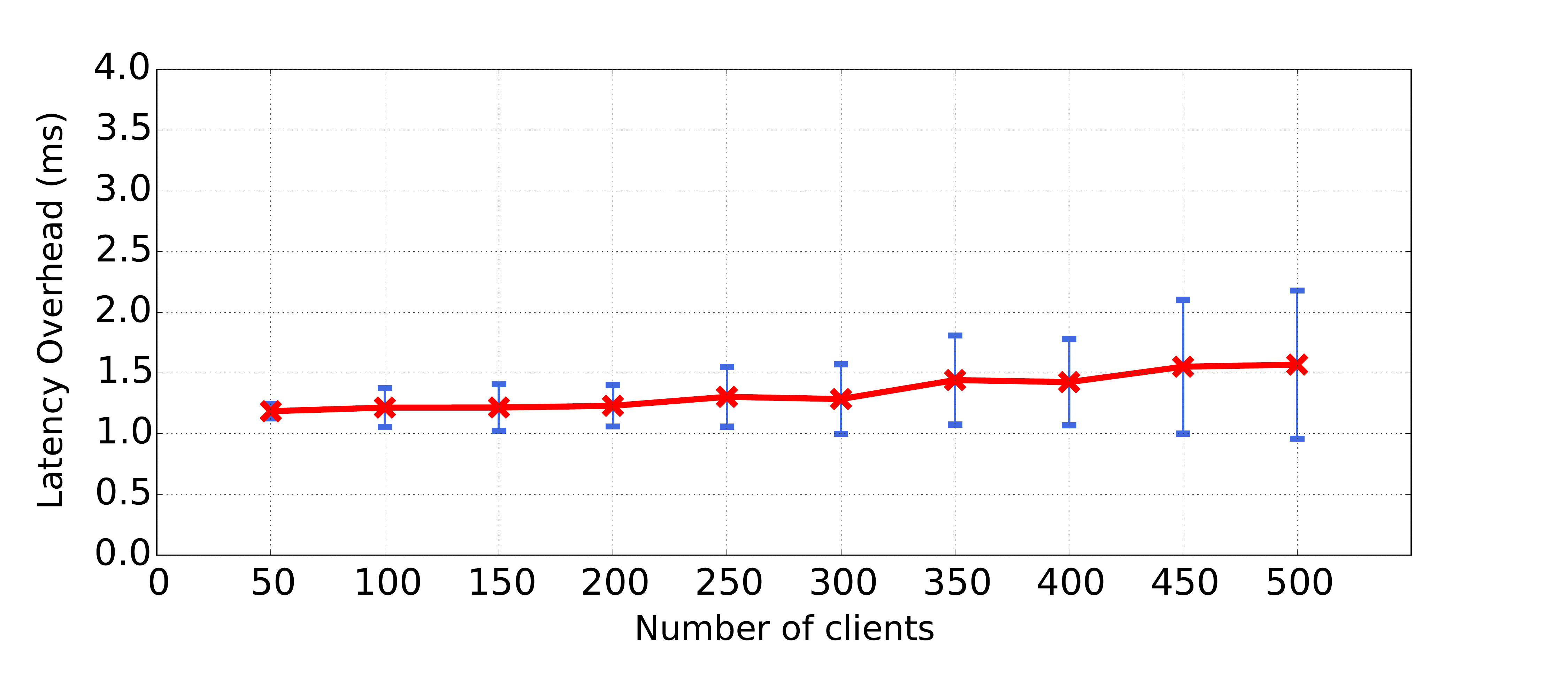}
\caption{Latency overhead of the system where $50$ to $500$ users simultaneously send traffic at rates $\lambda_{P} = \lambda_{L} = \lambda_{D} = 10$ per minute and mix nodes generate loop cover traffic at rate $\lambda_{M} = 10$ per minute. We assume that there is not additional delay added to the messages by the senders.}
\label{latency_mix}
\end{figure}

\paragraph{Latency Overhead \& Scalability.}
End-to-end latency overhead is the cost of routing and decoding relayed messages, without any additional artificial delays. We run simulations to measure its sensitivity to the number of users participating in the system. 

We measure the time needed to process a single packet by a mix node, which is approximately $0.6\,ms$. This cost is dominated by the scalar multiplication of an elliptic curve point and symmetric cryptographic operations. For the end-to-end measurement, we run \sysname with a setup where all users have the same rates of sending real and cover messages, such that $\lambda_{P} = \lambda_{D} = \lambda_{L} = 10$ messages per minute and mix servers generate loops at rate $\lambda_{M} = 10$ messages per minute. All clients set a delay of $0.0$ seconds for all the hops of their messages -- to ensure we only measure the system overhead, not the artificial mixing delay.

\Cref{latency_mix} shows that increasing the number of online clients, from $50$ to $500$, raises the latency overhead by only $0.37\,ms$. The variance of the processing delay increases with the amount of traffic in the network, but more clients do not significantly influence the average latency overhead. Neither the computational power of clients nor mix servers nor the communication between them seem to become bottlenecks for these rates. Those results show that the increasing number of users in the network does not lead to any bottleneck for our parameters. The measurements presented here are for a network of $6$ mix nodes, however we can increase the system capacity by adding more servers. Thus, \sysname scales well for an increasing number of users.

\begin{figure}[t!]
\centering
\includegraphics[width=\columnwidth]{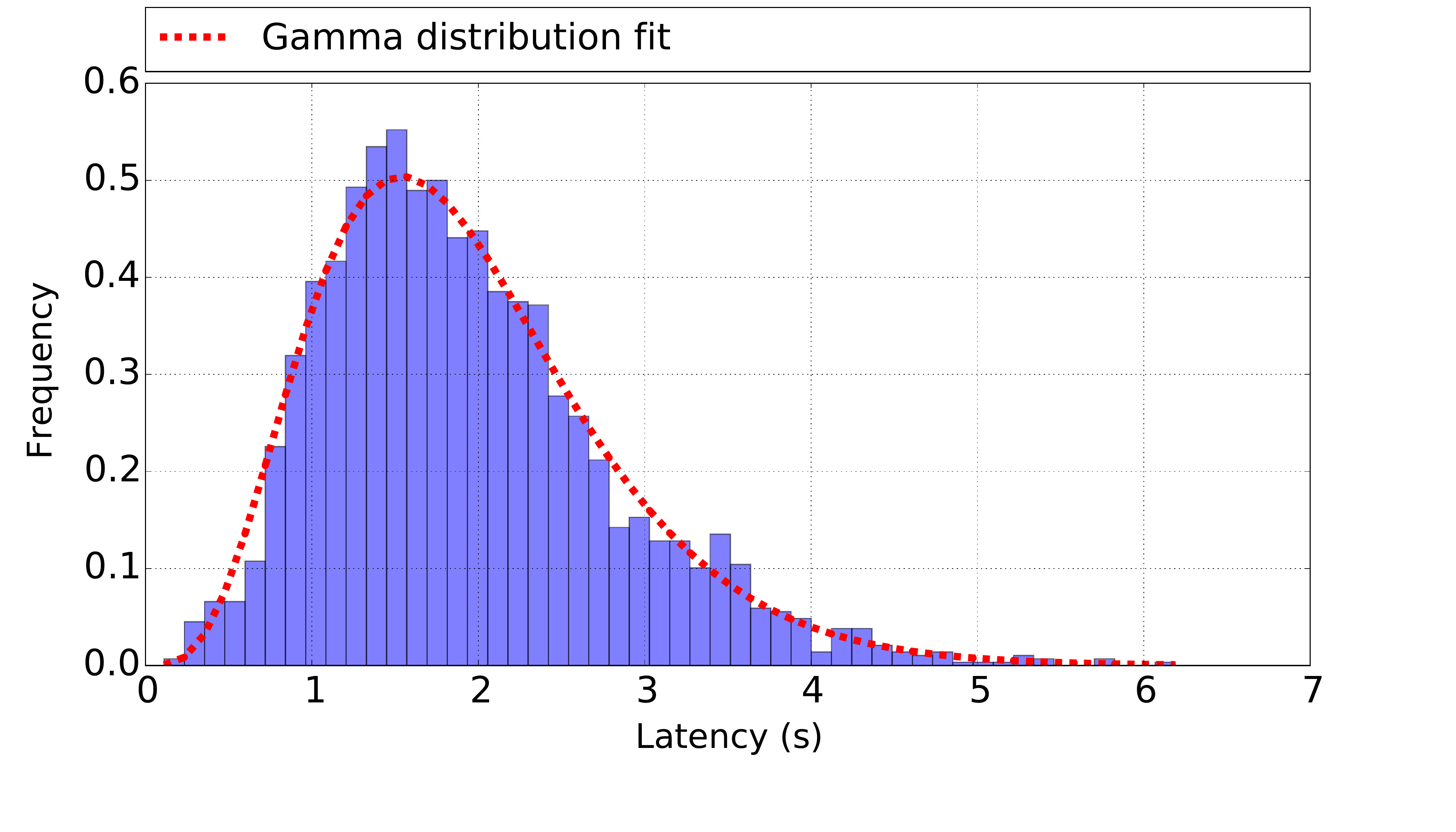}
\caption{End-to-end latency histogram measured through timing mix node loops. We run $500$ users actively communicating via \sysname at rates $\lambda_{P} = \lambda_{L} = \lambda_{D} = 60$ per minute and  $\lambda_{M} = 60$ per minute. The delay for each hop is drawn from $Exp(2)$. The latency of the message is determined by the assigned delay and fits the Gamma distribution with mean $1.93$ and standard deviation $0.87$.}
\label{latency_histogram}
\end{figure}

We also investigate how increasing the delays through Poisson Mixing with $\mu=2$ affects the end-to-end latency of messages. We measure this latency through timing mix heartbeat messages traversing the system. \Cref{latency_histogram} illustrates that when the mean delay $1/\mu$ sec. is higher than the processing time ($\sim 1\,ms-2\,ms$), the end-to-end latency is determined by this delay, and follows the Gamma distribution with parameter being the sum of the exponential distribution parameter over the number of servers on the path. The good fit to a gamma distribution provides evidence that the implementation of \sysname is faithful to the queuing theory models our analysis assumes.





\section{Related Work}\label{sec:related_works}


All anonymous communication designs share the common goal of hiding users' communication patterns from adversaries. 
Simultaneously minimizing latency and communication overhead while still providing high anonymity 
is challenging. We survey other anonymous systems and compare them with \sysname (a summary is provided in \Cref{tab:related_work_comparison}).


\paragraph{Early designs.} 
Designs based on Chaum's mixes~\cite{Chaum81} can support both high and low latency communication; all sharing the basic principles of mixing and layered encryption.
Mixmaster~\cite{moller2003mixmaster} supports sender anonymity using messages encryption 
but does not ensure receiver anonymity. Mixminion~\cite{danezis2003mixminion}
uses fixed sized messages
and supports anonymous replies
and ensures forward anonymity using link encryption between nodes. 
As a defense against traffic analysis, but at the cost of high-latencies, both designs delay incoming messages by collecting them in a pool that is flushed every $t$ seconds (if a fixed message threshold is reached).


In contrast, Onion routing~\cite{goldschlag1999onion} was developed for low-latency anonymous communication.
Similar to mix designs, each packet is encrypted in layers, and is decrypted by a chain of authorized onion routers.
Tor~\cite{dingledine2004tor}, the most popular low-latency anonymity system, is an overlay network of onion routers.   
Tor protects against sender-receiver message linking against a partially global adversary and ensures perfect forward secrecy, 
integrity of the messages, and congestion control. However, Tor is vulnerable to traffic analysis attacks, if an adversary can observe 
the ingress and egress points of the network.
 

\begin{table*}[htp]
  \centering
  \setlength{\arrayrulewidth}{.2em}
\resizebox{\textwidth}{!}{
 \begin{tabular}{lcccccccc} 
   & \textbf{Low}     &           \textbf{Low Communication}          &          \textbf{Scalable}              &  \textbf{Asynchronous} & \textbf{Active}  
   & \textbf{Offline} & \textbf{Resistance} \\ 
   & \textbf{Latency} &               \textbf{Overhead} &     \textbf{Deployment}        & \textbf{Messaging$\dagger$} & \textbf{Attack Resistant} 
   & \textbf{Storage*} & \textbf{to GPA} \\ 
 \hline
 \addlinespace
 \textbf{Loopix} & \Checkmark & \Checkmark & \Checkmark & \Checkmark & \Checkmark & \Checkmark & \Checkmark \\ 
 \addlinespace
 \textbf{Dissent}~\cite{wolinsky2012dissent} & \Cross & \Cross & \Cross & \Cross & \Checkmark & \Cross & \Checkmark \\
 \addlinespace
 \textbf{Vuvuzela}~\cite{van2015vuvuzela} & \Cross & \Cross & \Checkmark & \Cross & \Checkmark & \Cross & \Checkmark \\
 \addlinespace
 \textbf{Stadium}~\cite{tyagi2016stadium} & \Cross & \Checkmark & \Checkmark & \Cross & \Checkmark & \Cross & \Checkmark \\
 \addlinespace
 \textbf{Riposte}~\cite{corrigan2015riposte} & \Cross & \Cross & \Checkmark & \Cross & \Checkmark & \Cross & \Checkmark \\ 
 \addlinespace
 \textbf{Atom}~\cite{kwon2016atom} & \Cross & \Checkmark & \Checkmark & \Cross & \Checkmark & \Cross & \Checkmark \\ 
 \addlinespace
 \textbf{Riffle}~\cite{kwon2015riffle} & \Checkmark & \Checkmark & \Cross & \Cross & \Checkmark & \Cross & \Checkmark \\ 
 \addlinespace
 \textbf{AnonPoP}~\cite{gelernter2016anonpop} & \Cross & \Checkmark & \Checkmark & \Cross & \Cross &  \Checkmark & \Checkmark \\ 
 \addlinespace
 \textbf{Tor}~\cite{dingledine2004tor} & \Checkmark & \Checkmark & \Checkmark & \Checkmark & \Cross & \Cross & \Cross \\ 
\end{tabular}
}
 \caption{Comparison of popular anonymous communication systems. By *, we mean if the
    design intentionally incorporates provisions for delivery of messages when a user is offline, perhaps for a long period
of time. By $\dagger$, we mean that the system operates continuously and does not depend on synchronized rounds for its security properties and users do not need to coordinate to communicate together.
}\label{tab:related_work_comparison}
\end{table*}

\phead{Recent designs}
Vuvuzela~\cite{van2015vuvuzela} protects against both passive and active adversaries as long as there is one honest mix node. 
Since Vuvuzela operates in rounds, offline users lose the ability to 
receive messages and all messages must traverse a single chain of relay servers. 
\sysname does not operate in rounds, allows off-line users to receive messages and uses parallel mix nodes to improve the scalability of the network. 

Stadium~\cite{tyagi2016stadium} and AnonPop~\cite{gelernter2016anonpop} refine Vuvuzela; both operating in rounds making the routing of messages dependent on the dynamics of others.
Stadium is scalable, but it lacks offline storage, whereas AnonPop does provide offline message storage.
\sysname also provides both properties, and because it operates continuously avoids user synchronization issues.  


Riposte~\cite{corrigan2015riposte} is based on a \emph{write} PIR scheme in which users write their 
messages into a database, without revealing the row into which they wrote to the database server. 
Riposte enjoys low communication-overhead and 
protects against traffic analysis and denial of service attacks, but requires long 
epochs and a small number of clients writing into the database simultaneously.
In contrast to \sysname, it is suitable for high-latency applications.

Dissent~\cite{chaum1988dining}, based on DC-networks~\cite{chaum1988dining}, offers resilience against a 
GPA and some active attacks, but at significantly higher delays and scales to only several thousand clients.


Riffle~\cite{kwon2015riffle} introduces a new verifiable shuffle technique to achieve sender anonymity. 
Using PIR, Riffle guarantees receiver anonymity in the presence of an active adversary, as well as both sender and receiver anonymity, but it cannot support a large user base. 
Riffle also utilizes rounds protect traffic analysis attacks. Riffle is not designed for Internet-scale anonymous communication, like \sysname, but for supporting intra-group communication.


Finally, Atom~\cite{kwon2016atom} combines a number of novel techniques to
provide mid-latency communication, strong protection against passive adversaries and uses zero knowledge proofs between servers to resist active attacks. 
Performance scales horizontally, however latency comparisons between \sysname and Atom are difficult due to the dependence on pre-computation in Atom. 
Unlike \sysname, 
Atom is designed for latency tolerant uni-directional anonymous communication applications with only sender anonymity in mind.

\section{Discussion \& Future Work}\label{sec:discussion}

As shown in \Cref{sec:system_analysis}, the security of \sysname heavily depends on the ratio 
of the rate of traffic sent through the network and the mean delay at every mix node. Optimization of this ratio application dependent. For applications with small number of messages and delay tolerance, a small amount of cover traffic can guarantee security.

\sysname achieves its stated security and performance goals. However, there are many other facets of the design space that have been left for future work. For instance, reliable messages delivery, session management, and flow control while all avoiding inherent risks, such as statistical disclosure attacks~\cite{danezis2003statistical}, are all fruitful avenues of pursuit.

We also leave the analysis of replies to messages as future work. \sysname currently allows two methods if the receiver does not already know the sender a priori: 
we either attach the address of the sender to each message payload, or provide a single-use anonymous reply block~\cite{danezis2003mixminion,danezis2009sphinx}, which enables different use-cases.

 
The \sysname architecture deliberately relies on established providers to connect to and authenticate end-users. 
This architecture brings a number of potential benefits, such as resistance to Sybil attacks, enabling anonymous 
blacklisting~\cite{henry2013thinking} and payment gateways~\cite{AndroulakiRSSB08} to mitigate flooding attacks and other abuses of the system, and privacy preserving measurements~\cite{elahi2014privex,JansenJ16} about client and network trends and the security stance of the system. All of this analysis is left for future work. 



It is also apparent that an efficient and secure private lookup system, one that can deliver network state and keying information to its users, is necessary to support modern anonymous communications. Proposals of stand-alone `presence' systems such as DP5~\cite{borisov2015dp5} and MP3~\cite{parhi2016mp3} provide efficient lookup
methods, however, we anticipate that tight integration between the lookup and anonymity systems may bring mutual performance and security benefits, which is another avenue for future work.

\section{Conclusion}\label{sec:conclusion}


The \sysname mix system explores the design space frontiers of low-latency mixing. We balance cover traffic and message delays to achieve a tunable trade-off between real traffic and cover traffic, and between latency and good anonymity.
Low-latency incentivizes early adopters to use the system, as they benefit from good performance. Moreover, the cover traffic introduced by both clients and mix servers provides security in the presence of a smaller user-base size. 
In turn this promotes growth in the user-base leading on one hand to greater security~\cite{dingledine2006anonymity}, and on the other a tuning down of cover traffic over time. 

\sysname is the first system to combine a number of best-of-breed techniques: we provide definitions inspired by AnoA~\cite{backes2013anoa} for our security properties; improve the analysis of simplified variants of stop-and-go-mixing as a Poisson mix~\cite{kesdogan1998stop}; we use restricted topologies to promote good mixing~\cite{dingledine2004synchronous}; we deploy modern active attack 
mitigations based on loops~\cite{danezis2003heartbeat}; and we use modified modern cryptographic packet formats, such as 
Sphinx~\cite{danezis2009sphinx}, for low information leakage. Our design, security and performance analysis, and empirical evaluation shows they work well together to provide strong security guarantees.

The result of composing these different techniques -- previously explored as separate and abstract design options -- is a design that is strong against global network level adversaries without the very high-latencies traditionally associated with mix systems~\cite{moller2003mixmaster,danezis2003mixminion}. Thus, \sysname revitalizes message-based mix systems and makes them competitive once more against onion routing~\cite{goldschlag1999onion} based solutions that have dominated the field of anonymity research since Tor~\cite{dingledine2004tor} was proposed in 2004.

\ifanonymous
\else
\paragraph{Acknowledgments} We thank Claudia Diaz and Mary Maller for the helpful discussions. In memory of Len Sassaman.
This work was supported by NSERC through a Postdoctoral Fellowship Award, the Research Council KU Leuven: C16/15/058, the European Commission through H2020-DS-2014-653497 PANORAMIX, the EPSRC Grant EP/M013-286/1, and the UK Government Communications Headquarters (GCHQ), as part of University College London's status as a recognised Academic Centre of Excellence in Cyber Security Research.

\fi

\bibliographystyle{acm}
\bibliography{sections/references}

\appendix

\section{Appendix}\label{sec:appendix}
\subsection{Incremental Computation of the Entropy Metric}
Let $X$ be a discrete random variable over the finite set $\mathcal X$ with probability mass function $p(x) = \Pr(X = x)$. The Shannon entropy $H(X)$~\cite{shannon2001mathematical} of a discrete random variable $X$ is defined as 
\begin{align}
	H(X) = - \sum_{x \in \mathcal{X}} p(x)\log p(x).
\end{align}

Let $o_{n,k,l}$ be an observation as defined in \Cref{sec:poisson_sec} for a pool at time $t$. We note that any outgoing message will have a distribution over being linked with past input messages, and the entropy $H_t$ of this distribution is our anonymity metric. $H_t$ can be computed incrementally given the size of the pool $l$ (from previous mix rounds) and the entropy $H_{t-1}$ of the messages in this previous pool, and the number of messages $k$ received since a message was last sent:
\begin{align}
\begin{split}
	H_t = &H\left( \left\{ \frac{k}{k+l}, \frac{l}{k+l} \right\} \right) \\
	    & + \frac{k}{k+l} \log k 
	     + \frac{l}{k+l} H_{t-1},
\end{split}
\end{align} 
for any $t > 0$ and $H_0 = 0$. Thus for sequential observations we can incrementally compute the entropy metric for each outgoing message, without remembering the full history of the arrivals and departures at the Poisson mix. We use this method to compute the entropy metric illustrated in \Cref{fig:entropy}.

\subsection{Proof of \Cref{probability2}}

Let us assume, that in mix node $M_{i}$ there are $n'$ messages at a given moment, among which is a target message $m_{t}$. Each message has a delay $d_{i}$ drawn from the exponential distribution with parameter $\mu$. The mix node generates loops with distribution $\pois(\lambda_{M})$. The adversary observes an outgoing message $m$ and wants to quantify whether this outgoing message is her target message. The adversary knows, that the output of the mix node can be either one of the messages inside the mix or its loop cover message. 
Thus, for any message $m_t$, the following holds
\begin{align}\label{eq1}
\begin{split}
	&\pr{m = m_{t}} = \pr{m \neq loop} \cdot \pr{m = m_{t} | m \neq loop} \\
\end{split}
\end{align}
We note that the next message $m$ is a loop if and only if the next loop message is sent before any of the messages within the mix, i.e., if the sampled time for the next loop message is smaller than any of the remaining delays of all messages within the mix. We now leverage the memoryless property of the exponential distribution to model the remaining delays of all $n'$ messages in the mix as fresh random samples from the same exponential distribution.
\begin{align}\label{eq-noloop}
\begin{split}
	&\pr{m \neq loop} =  1 - \pr{m = loop} \\
	& = 1 - \pr{X < d_{1} \land X < d_{2} \land \ldots \land X < d_{n'} }\\
	& =  1 - \pr{X < min\{d_{1}, d_{2}, \ldots d_{n'}\} }\\
\end{split}
\end{align}

We know, that $d_{i} \sim Exp(\mu) \text{ for all } i \in \set{1, \ldots, n'}$ and $X \sim Exp(\lambda_{M})$. Moreover, we know that 
the minimum of $n$ independent exponential random variables with rate $\mu$ is an exponential random variable with parameter $\sum_{i}^{n'} \mu$.
Since all the delays $d_{i}$ are independent exponential variables with the same parameter, we have for $Y = \min\{d_{1}, d_{2}, \ldots d_{n'}\}$, $Y \sim Exp(n'\mu)$.
Thus, we obtain the following continuation of \Cref{eq-noloop}.
\begin{align}\label{eq-noloopprob}
\begin{split}
	&\pr{m \neq loop} = 1 - \pr{X < Y} \\
	& =  1 - \int_{0}^{\infty} \pr{X < Y | X = x}\pr{X = x} dx  \\
	& =  1 - \int_{0}^{\infty} \pr{x < Y}\lambda_{M}e^{-\lambda_{M}x} dx \\
	& =  1 - \int_{0}^{\infty} e^{-n'\mu x}\lambda_{M}e^{-\lambda_{M}x} dx \\
	& =  1 - \frac{\lambda_{M}}{\lambda_{M} + n\mu}\\
	& =  \frac{n'\mu}{n'\mu + \lambda_{M}}
\end{split}
\end{align} 
Since the probability to send a loop depends only on the \emph{number} of messages in a mix, but not on which messages are in the mix, this probability is independent of the probability from \Cref{probability1}. \Cref{probability2} follows directly by combining \Cref{probability1} and \Cref{eq-noloopprob}, with $n' = k+l$. We get for messages $m_1$ that previously were in the mix,
\[
\begin{aligned}
	\pr{m = m_{1}} &= \pr{m \neq loop} \cdot \pr{m = m_{1} | m \neq loop} \\
	&=  \frac{(k+l) \mu}{(k+l)\mu + \lambda_{M}} \cdot \frac{k}{n(k+l)} \\
	&=  \frac{k}{n} \cdot \frac{\mu}{(k+l)\mu + \lambda_{M}}. \\
\end{aligned}
\]
Analogously, we get for $m_2$,
\[
\begin{aligned}
	\pr{m = m_{2}} &= \pr{m \neq loop} \cdot \pr{m = m_{2} | m \neq loop} \\
	&=  \frac{(k+l) \mu}{(k+l)\mu + \lambda_{M}} \cdot \frac{1}{k+l} \\
	&=  \frac{\mu}{(k+l)\mu + \lambda_{M}}.\\
\end{aligned}
\]
This concludes the proof.

%



\end{document}